\documentclass[12pt,a4paper]{article}
\usepackage{amsmath,caption}
\usepackage[top=1.2in, bottom=1.2in, left=1.1in, right=1.1in]{geometry}

\usepackage[round]{natbib}
\usepackage{color,soul}

\DeclareCaptionStyle{italic}[justification=centering]
 {labelfont={bf},textfont={it},labelsep=colon}
\usepackage{graphicx,psfrag,epsf,textcomp}
\usepackage{enumerate}
\usepackage{natbib}
\usepackage{url,xcolor} 
\usepackage{booktabs, subfig, bm, paralist,mathpazo,tikz,todonotes,longtable,microtype}

\usepackage[pdftex,colorlinks=true]{hyperref}
\definecolor{darkblue}{rgb}{0,0,.6}
\hypersetup{citecolor=darkblue,linkcolor=darkblue,urlcolor=darkblue}

\newcommand{\blind}{0}

\graphicspath{{plots/}}

\newsavebox\CBox
\def\textBF#1{\sbox\CBox{#1}\resizebox{\wd\CBox}{\ht\CBox}{\textbf{#1}}}

\definecolor{a0}{rgb}{0.0, 0.5, 0.0}
\definecolor{bistre}{rgb}{0.24, 0.17, 0.12}
\definecolor{amethyst}{rgb}{0.6, 0.4, 0.8}
\definecolor{blue-violet}{rgb}{0.54, 0.17, 0.89}
\definecolor{Rcolor}{RGB}{150,160,190}
\definecolor{blush}{rgb}{0.87, 0.36, 0.51}
\definecolor{brightturquoise}{rgb}{0.03, 0.91, 0.87}
\definecolor{burntorange}{rgb}{0.8, 0.33, 0.0}

\begin{document}

\def\spacingset#1{\renewcommand{\baselinestretch}%
{#1}\small\normalsize} \spacingset{1}

\if0\blind
{
  \title{\bf Bayesian bandwidth estimation and semi-metric selection for a functional partial linear model with unknown error density}
  \author{
    Han Lin Shang\footnote{Postal address: Research School of Finance, Actuarial Studies and Statistics, Level 4, Building 26C, Kingsley Street, Australian National University, Acton ACT 2601, Canberra, Australia; Telephone number: +61(2) 612 50535; Fax number: +61(2) 612 50087; Email: hanlin.shang@anu.edu.au} \\
 Australian National University}
  \maketitle
} \fi

\if1\blind
{
  \bigskip
  \bigskip
  \bigskip
  \begin{center}
    {\LARGE\bf Bayesian bandwidth estimation and semi-metric selection for a functional partial linear model with unknown error density}
\end{center}
  \medskip
} \fi

\bigskip
\begin{abstract}
This study examines the optimal selections of bandwidth and semi-metric for a functional partial linear model. Our proposed method begins by estimating the unknown error density using a kernel density estimator of residuals, where the regression function, consisting of parametric and nonparametric components, can be estimated by functional principal component and functional Nadayara-Watson estimators. The estimation accuracy of the regression function and error density crucially depends on the optimal estimations of bandwidth and semi-metric. A Bayesian method is utilized to simultaneously estimate the bandwidths in the regression function and kernel error density by minimizing the Kullback-Leibler divergence. For estimating the regression function and error density, a series of simulation studies demonstrate that the functional partial linear model gives improved estimation and forecast accuracies compared with the functional principal component regression and functional nonparametric regression. Using a spectroscopy dataset, the functional partial linear model yields better forecast accuracy than some commonly used functional regression models. As a by-product of the Bayesian method, a pointwise prediction interval can be obtained, and marginal likelihood can be used to select the optimal semi-metric. 

\vspace{.2in}

\end{abstract}

\newpage
\spacingset{1.45}

\section{Introduction}

In scalar-on-function regression, parametric regression models are useful for interpreting the linear relationship between scalar response and functional predictor, while nonparametric regression models may capture a possible nonlinear relationship, and thus may improve the estimation and prediction accuracies of the regression function. By combining the advantages of parametric and nonparametric regression models, the semiparametric regression models, such as the functional partial linear model, have received increasing attention in the literature \citep[e.g.,][]{Lian11}.

Despite rapid developments in the estimations of functional partial linear models, the optimal selections of semi-metric and bandwidth remain largely unexplored. To address this, we consider the optimal parameter selections from the perspective of error-density estimation in functional regression. The estimation of error density is important for understanding the residual behavior of regression models and assessing the adequacy of the error distribution assumption \citep[e.g.,][]{AV01, CS08}. The estimation of error density is also useful for testing the symmetry of the residual distribution \citep[e.g.,][]{AL97, ND07}; for the estimation of the density of response variable \citep[e.g.,][]{EJ12}; and is important for statistical inference, prediction and model validation \citep[e.g.,][]{Efromovich05, MN10}.  

In a nonparametric scalar-on-function regression, \cite{Shang13} proposes a Bayesian bandwidth estimation method for determining the bandwidth of the Nadaraya-Watson (NW) estimator for estimating regression mean function and the bandwidth of the kernel-form error density. Also, in nonparametric scalar-on-function regression, \cite{Shang16} proposes a Bayesian bandwidth estimation method for determining the bandwidth of the NW estimator for estimating regression quantile function and the bandwidth of the kernel-form error density. Further, \cite{Shang16} uses marginal likelihood as a means of selecting the optimal semi-metric. Building on the early work by \cite{ZKS11}, \cite{ZK11} and \cite{Shang13, Shang14, Shang13b, Shang16, Shang19}, we consider a kernel error-density estimator which explores data-driven features, such as asymmetry, skewness, and multi-modality, and relies on residuals obtained from the estimated regression function and bandwidth of residuals. Differing from those early work, we derive an approximate likelihood and a posterior for the functional partial linear model (a semiparametric model). We present a Markov chain Monte Carlo (MCMC) sampling algorithm for simultaneously sampling bandwidth parameters, linear regression coefficient function $\beta(t)$, and nonlinear NW estimator of $m(z)$. Through a series of simulation studies, the functional partial linear model gives improved estimation and forecast accuracies of the regression mean function compared with the functional principal component regression and nonparametric scalar-on-function regression.

While the selection of bandwidth determines the amount of smoothing, the selection of semi-metric plays an important role in measuring distances among functions. In nonparametric functional data analysis, the optimal bandwidth can be determined either by functional cross-validation \citep{RV07} or by a Bayesian method incorporating the information on error density \citep{Shang13}. However, the optimal selection of semi-metric is rather arbitrary. As noted by \citet[][Chapters 3 and 13]{FV06}, a semi-metric based on a derivative should be used for a set of smooth functional data; a semi-metric based on a dimension-reduction technique, such as functional principal component analysis (FPCA), should be used for a set of rough functional data. To the best of our knowledge, little attention has been paid for selecting which semi-metric is adequate based on rigorous statistical criteria.

This lack of rigorous statistical criteria motivates our investigation of a Bayesian method for estimating bandwidths in the regression function and error density simultaneously, and for selecting the optimal semi-metric, based on the notion of marginal likelihood (also referred to \cite{KR95} as the evidence). The marginal likelihood reflects a summary of evidence provided by the data supporting a choice of semi-metric as opposed to its competing semi-metric. The optimal semi-metric is that which has the largest marginal likelihood. With the marginal likelihood, it is straightforward to calculate the Bayes factor for measuring the strength of evidence between any two semi-metrics \citep[e.g.,][]{KR95}. Selecting the optimal semi-metric and bandwidths often improves the estimation and prediction accuracies of the regression function, including the functional partial linear model considered here. 

The remainder of this paper is organized as follows. In Section~\ref{sec:2}, we introduce the partial linear model with the functional covariate, previously studied by \cite{Lian11}, and the estimation procedure in \cite{AV15}, \cite{AFV15} and \cite{NAV17}. In Section~\ref{sec:3}, we review the Bayesian bandwidth-estimation method and introduce the notion of marginal likelihood computed as a by-product of the MCMC. The optimal semi-metric is determined by having the largest marginal likelihood among several possible semi-metrics. Using a series of simulation studies in Section~\ref{sec:4}, we evaluate and compare the estimation accuracies of the regression function and error density. Also, we compare the forecast accuracy of regression function among the functional partial linear model, functional principal component regression, and functional nonparametric regression. In Section~\ref{sec:5}, we apply the proposed method to a spectroscopy dataset in the food quality control and compare its forecast accuracy with 13 commonly used functional regression models. Conclusions are presented in Section~\ref{sec:6}, along with some reflections on how the method presented here could be extended.

\section{Model and estimator}\label{sec:2}

There is an increasing amount of literature on the development of nonparametric functional estimators, such as the functional NW estimator \citep{FV06}, the functional local-linear estimator \citep{BEM11}, the functional $k$-nearest neighbor estimator \citep{BFV09}, and the distance-based local-linear estimator \citep{BDF10}. In the functional partial linear model, we choose to estimate the \textit{conditional mean} by the functional NW estimator because of its simplicity and robustness against a large gap in design points.

\subsection{A functional partial linear model}
  
When modeling scalar responses and functional predictors, some functional variables are related to responses linearly while other variables have nonlinear relationships with the response. We consider a random data triplet $(\mathcal{X}, \mathcal{Z}, y)$, where $y$ is a real-valued response and the functional random variable $\mathcal{X}$ is valued in Hilbert space containing square-integrable functions and $\mathcal{Z}$ is valued in some infinite-dimensional semi-metric vector space $(\mathcal{F}, d(\cdot,\cdot))$. Let $(\mathcal{X}_i, \mathcal{Z}_i, y_i)_{i=1,2,\dots,n}$ be a sample of data triplets that are independent and identically distributed (iid) as $(\mathcal{X}, \mathcal{Z}, y)$. We consider a functional partial linear model with homoscedastic errors. Given a set of observations $(\mathcal{X}_i, \mathcal{Z}_i, y_i)$, the regression model can be expressed as
\begin{equation*}
y_i = \int_{t\in\mathcal{I}} \mathcal{X}_i(t)\beta(t)dt + m(\mathcal{Z}_i) + \varepsilon_i,
\end{equation*}
where $\mathcal{I}$ represents function support range, $\beta(\cdot)$ is a regression coefficient function, $m(\cdot)$ is an unknown smooth real function that captures the possible nonlinear relationship between $\mathcal{Z}$ and $y$ and $(\varepsilon_1,\dots,\varepsilon_n)$ are iid random error satisfying
\begin{equation*}
\text{E}\left(\varepsilon_i|\mathcal{X}_i, \mathcal{Z}_i\right) = 0.
\end{equation*}

Following the estimation procedure of \cite{AV06}, it is natural to estimate the regression coefficient function $\beta(\cdot)$ and the smooth function $m(\cdot)$ by the ordinary least squares and functional NW estimators, given by
\begin{align}
\widehat{\beta}_h(t) &= \left[\bm{\widetilde{\mathcal{X}}}^{\top}_h(t)\bm{\widetilde{\mathcal{X}}}_h(t)\right]^{+}\bm{\widetilde{\mathcal{X}}}_h^{\top}(t)\bm{\widetilde{y}}_h,\label{eq:3} \\
\widehat{m}_h(z) &= \sum_{i=1}^nw_h(z,\mathcal{Z}_i)\left[y_i-\langle\mathcal{X}_i(t), \widehat{\beta}_h(t)\rangle\right],\label{eq:4}
\end{align}
where $^{+}$ represents generalized inverse and $^{\top}$ represents the column vector; $\bm{\mathcal{X}} = \left(\mathcal{X}_1,\dots,\mathcal{X}_n\right)^{\top}$ and $\bm{y}=\left(y_1,\dots,y_n\right)^{\top}$, $\bm{\widetilde{\mathcal{X}}}_h=(\bm{I}-\bm{W}_{h})\bm{\mathcal{X}}$ and $\widetilde{\bm{y}}_h = (\bm{I}-\bm{W}_h)\bm{y}$, $\bm{W}_h = \left[w_{h}(\mathcal{Z}_i, \mathcal{Z}_j)\right]_{i,j}$ is a weight matrix with $w_h\big(z,\mathcal{Z}_i\big)$ being the NW-type weights,
\begin{equation*}
w_h(z,\mathcal{Z}_i) = \frac{K\left[d(z,\mathcal{Z}_i)/h\right]}{\sum^n_{j=1}K\left[d(z,\mathcal{Z}_j)/h\right]},
\end{equation*}
where $K(\cdot)$ is a kernel function, such as the Gaussian kernel function considered in this paper, $h$ is a positive real-valued estimated bandwidth parameter, controlling the trade-off between squared bias and variance in the mean squared error (MSE) and $d(\cdot,\cdot)$ is a semi-metric used to quantify differences among curves. 

The regression coefficient function $\widehat{\beta}_h(t)$ in~\eqref{eq:3} can be seen as the ordinary least squares estimator obtained by regressing the partial residual vector $\widetilde{\bm{y}}_h$ on the partial residual vector of functions $\widetilde{\bm{\mathcal{X}}}_h$. To avoid possible non-singularity, it is common to use a dimension-reduction technique, such as functional principal component regression  \citep[e.g.,][]{RO07, RGS+17, Lian11}. However, $\widehat{m}_h(z)$ in~\eqref{eq:4} can be seen as the functional NW estimator, with the partial residual vector as the response variable. Therefore, the estimation accuracy of $\widehat{\beta}_h(t)$ and $\widehat{m}_h(z)$ crucially depends on the optimal estimation of $h$. With an estimated bandwidth of the NW estimator, we obtain an estimate of $\beta(t)$ at each iteration of the MCMC sampling. Then, we take average over all iterations to obtain the final estimates, namely $\widehat{h}$, $\widehat{\beta}_h(t)$ and $\widehat{m}_h(z)$.

In functional nonparametric regression, the bandwidth parameter is commonly determined by functional cross-validation \citep[e.g.,][]{FV09, BFV09, BFV10, FHV10}. Functional cross-validation aims to select a bandwidth that minimizes the squared loss function and has the appealing feature that no estimation of the error variance is required. However, as stated, an accurate estimation of error density is important. Thus, we consider a Bayesian method to estimate bandwidth parameters in the regression function and error density simultaneously. The Bayesian method aims to select a bandwidth that minimizes the Kullback-Leibler (KL) divergence.

\subsection{Choice of semi-metric}

The choice of semi-metric has effects on the size and form of neighborhoods and can thus control the concentration properties of functional data. Although \citet[][p.193]{FV06} note that the selection of the semi-metric remains an open question, there is no method to quantify which semi-metric is adequate, or how to select the optimal semi-metric, using statistical criteria.

From a practical perspective, the semi-metric based on derivative should be utilized for a set of smooth functional data. This semi-metric can be expressed as
\begin{equation*}
d_q^{\text{deriv}}(\mathcal{Z}_i,\mathcal{Z}) = \sqrt{\int_t\left[\mathcal{Z}_i^{(q)}(t) - \mathcal{Z}^{(q)}(t)\right]^2dt},
\end{equation*}
where $\mathcal{Z}^{(q)}$ is the $q$\textsuperscript{th}-order derivative of $\mathcal{Z}$, in which first and second derivatives are commonly used in practice \citep[e.g.,][]{Goutis98, FV02, FV09}. Computationally, the semi-metric based on the $q$\textsuperscript{th}-order derivative uses a $B$-spline approximation for each curve, and a derivative of a $B$-spline function can be directly computed by differencing the $B$-spline coefficients and $B$-spline basis functions of one order lower \citep[for more detail, see][Section 3.4]{deBoor01, FV06}. 

For a set of rough functional data, a semi-metric based on FPCA is commonly used \citep[for detail on the choice of semi-metric from the practical and theoretical perspectives, see][Chapters 3 and 13]{FV06}. The semi-metric based on FPCA computes proximities among a set of rough curves using the principal component scores. FPCA reduces the functional data to a finite-dimensional space (i.e., $K$ number of components). The finite-dimensional representation can be expressed as
\begin{equation*}
d_K(\mathcal{Z}_i,\mathcal{Z}) \approx \sqrt{\sum^K_{\zeta=1}(\beta_{\zeta,i}-\beta_{\zeta})^2\|\phi_{\zeta}(t)\|^2},
\end{equation*}
where $K$ represents the number of retained functional principal components, and $\|v\|=\sqrt{\langle v,v\rangle}$ represents the induced norm. 

\section{Bayesian method}\label{sec:3}

In Section~\ref{sec:3.1}, we briefly review \citeauthor{Shang13}'s \citeyearpar{Shang13} Bayesian method for selecting optimal bandwidth. As a by-product of the MCMC, the optimal semi-metric can be determined with the largest marginal likelihood among a set of semi-metrics (as presented in Section~\ref{sec:3.2}).

\subsection{Bayesian bandwidth estimation}\label{sec:3.1}

\cite{ZKS11} consider a kernel-form error density in a nonparametric regression model, while \cite{ZK11} consider the kernel-form error density in a nonlinear parametric model where Bayesian sampling approaches are used. Following the early work by \cite{ZKS11} and \cite{ZK11}, the unknown error density $f(\varepsilon)$ can be approximated by a location-mixture Gaussian density, given by
\begin{equation*}
f(\varepsilon;b) = \frac{1}{n}\sum^n_{j=1}\frac{1}{b}\phi\left(\frac{\varepsilon-\varepsilon_j}{b}\right),
\end{equation*}
where $\phi(\cdot)$ is the probability density function of  the standard Gaussian distribution and Gaussian densities have means at $\varepsilon_j$, for $j=1,2,\dots,n$ and a common standard deviation $b$.

Given that errors are unknown in practice, we approximate them by residuals obtained from the functional principal component and functional NW estimators of the conditional mean. Given bandwidths $h$ and $b$, the kernel likelihood of $\bm{y} = (y_1,\dots,y_n)^{\top}$ is given by
\begin{equation}
\widehat{L}\left(\bm{y}|h,b\right) = \prod^n_{i=1}\Bigg[\frac{1}{n-1}\sum^n_{\substack{j=1\\ j\neq i}}\frac{1}{b}\phi\left(\frac{\widehat{\varepsilon}_{i}-\widehat{\varepsilon}_{j}}{b}\right)\Bigg],\label{eq:5}
\end{equation}
where $\widehat{\varepsilon}_i=y_i - \int_{t\in \mathcal{I}}\mathcal{X}_i(t)\widehat{\beta}(t)dt - \widehat{m}(\mathcal{Z}_i)$ represents $i$\textsuperscript{th} residual obtained from the estimated regression function. This likelihood is not proper since some terms are left out. Instead, the likelihood is a pseudo-likelihood. The pseudo-likelihood is a likelihood function associated with a family of probability distributions, which does not necessarily contain the true function \citep{GMT84}. As a consequence, the resulting Bayesian estimators, while consistent, may have an inaccurate posterior variance, and subsequent credible sets constructed by this posterior may not be asymptotically valid \citep[see also][]{VH03, Mueller13}.

By assuming that the independent priors of squared bandwidths $h^2$ and $b^2$ follow inverse Gamma distribution $\text{IG}\left(10^{-3},10^{-3}\right)$, the posterior density can be obtained by multiplying the kernel likelihood in~\eqref{eq:5} with the prior density. The posterior density can be expressed as (up to a normalizing constant)
\begin{equation*}
\pi\left(h^2,b^2|\bm{y}\right) \propto \widehat{L}\left(\bm{y}|h,b\right) \times \pi\left(h^2\right)\times \pi\left(b^2\right).\label{eq:posterior}
\end{equation*}

\subsection{Sampling algorithm}\label{sec:3.3}

Computationally, an adaptive random-walk Metropolis algorithm can be used for sampling both parameters $(h,b)$ iteratively \citep[see][]{GFS10}. The sampling procedure is briefly described as follows:
\begin{enumerate}
\item[1)] Specify a Gaussian proposal distribution and begin the sampling iteration process by choosing an arbitrary value of $(h^2, b^2)$ and denoting it as $(h_{(0)}^2, b_{(0)}^2)$. The starting points can be drawn from a uniform distribution $U(0,1)$.
\item[2)] From the initial bandwidths $h_{(0)}^2$ and $b_{(0)}^2$, we obtain an initial estimate of linear regression coefficient function, $\widehat{\beta}_{h_{(0)}}(t)$ from~\eqref{eq:3} and an initial estimate of nonlinear regression function, $\widehat{m}_{h_{(0)}}(z)$ from~\eqref{eq:4}.
\item[3)] At the $k$\textsuperscript{th} iteration, the current state $h_{(k)}^2$ is updated as $h_{(k)}^2=h_{(k-1)}^2+\tau_{(k-1)}u$, where $u$ is drawn from the proposal density which can be the standard Gaussian density and $\tau_{(k-1)}$ is an adaptive tuning parameter with an arbitrary initial value $\tau_{(0)}$. The updated $h_{(k)}^2$ is accepted with a probability given by
\begin{equation*}
\min\left\{\frac{\pi\left(h^2_{(k)},b^2_{(k-1)}\big|\bm{y}\right)}{\pi\left(h^2_{(k-1)},b^2_{(k-1)}\big|\bm{y}\right)},1\right\}.
\end{equation*}
\item[4)] The tuning parameter for the next iteration is set to
\[ \tau_{(k)} = \left\{ \begin{array}{ll}
\tau_{(k-1)}+c(1-\xi)/k & \mbox{\qquad if $h^2_{(k)}$ is accepted} \\
\tau_{(k-1)}-c\xi/k & \mbox{\qquad if $h^2_{(k)}$ is rejected} \end{array} \right.\]
where $c=\tau_{(k-1)}/(\xi-\xi^2)$ is a constant and $\xi$ is the optimal target acceptance probability, which is 0.44 for univariate updating \citep[e.g.,][]{RR09}.	
\item[5)] Conditional on $h_{(k)}^2$ and $\bm{y}$, we obtain an estimator of linear regression coefficient function, $\widehat{\beta}_{h_{(k)}}(t)$, and an NW estimator of nonlinear regression function, $\widehat{m}_{h_{(k)}}(z)$ \citep[see, e.g.,][]{CSZ17}.
\item[6)] Repeat Steps 2-5 for $b_{(k)}^2$, conditional on $h_{(k)}^2$ and $\bm{y}$.
\item[7)] Repeat Steps 2-6 for $M+N$ times, discard $\big(h^2_{(0)}, \beta_{h_{(0)}}(t), m_{h_{(0)}}(z), b^2_{(0)}\big)$, $\big(h^2_{(1)}, \beta_{h_{(1)}}(t), m_{h_{(1)}}(z), b^2_{(1)}\big)$, \\ $\dots$, $\left(h^2_{(M)}, \beta_{h_{(M)}}(t), m_{h_{(M)}}(z), b^2_{(M)}\right)$ for burn-in to let the effects of the transients wear off, estimate $\widehat{h}^2 = \frac{\sum^{M+N}_{k=M+1}h_{(k)}^2}{N}$, $\widehat{\beta}_{h}(t) = \frac{\sum^{M+N}_{k=M+1}\beta_{h_{(k)}}(t)}{N}$, $\widehat{m}_{h}(z) = \frac{\sum^{M+N}_{k=M+1}m_{h_{(k)}}(z)}{N}$, and $\widehat{b}^2=\frac{\sum^{M+N}_{k=M+1}b_{(k)}^2}{N}$. The burn-in period is taken to be $M=1,000$ iterations and the number of iterations after the burn-in is $N=10,000$ iterations.
\end{enumerate}

\subsection{Bayesian model selection}\label{sec:3.2}

In Bayesian inference, model selection is often conducted through the marginal likelihood of the model of interest against a competing model. The marginal likelihood is the expectation of likelihood with respect to the prior density of parameters and reflects a summary of evidence provided by the data supporting the model as opposed to its competing model. It is seldom calculated as the integral of the product of the likelihood and prior density of parameters, but instead, is often computed numerically \citep[e.g.,][]{GD94, Chib95}. We use the method proposed by \cite{Chib95} to compute the marginal likelihood.

Let $\bm{\theta}=\left(h,b\right)$ be the parameter vector and $\bm{y}=\left(y_1,\dots,y_n\right)$ be the data. \cite{Chib95} demonstrated that the marginal likelihood under model $A$ can be expressed as
\begin{equation*}
L_{\text{A}}(\bm{y}) = \frac{L_{\text{A}}(\bm{y}|\bm{\theta})\pi_{\text{A}}(\bm{\theta})}{\pi_{\text{A}}(\bm{\theta}|\bm{y})},
\end{equation*}
where $L_{\text{A}}(\bm{y}|\bm{\theta})$, $\pi_{\text{A}}(\bm{\theta})$ and $\pi_{\text{A}}(\bm{\theta}|\bm{y})$ denote the likelihood, prior and posterior, respectively. $L_{\text{A}}(\bm{y})$ is often computed at the posterior estimate of $\bm{\theta}$. The numerator has a closed form and can be computed analytically. The denominator can be estimated by its kernel-density estimator based on the simulated samples of $\bm{\theta}$ through a posterior sampler. Based on marginal likelihoods, the Bayes factor of model A against model B is defined as $L_{\text{A}}(\bm{y})/L_{\text{B}}(\bm{y})$, which can be used to decide whether model A or B is preferred, along with its degree of preference \citep{KR95}.

\subsection{Adaptive estimation of error density}

In the kernel-density estimation of directly observed data, it has been noted that the leave-one-out estimator is profoundly affected by extreme observations in the sample \citep[e.g.,][]{Bowman84}. When the actual error density has sufficient long tails, the leave-one-out kernel-density estimator with a global bandwidth estimated under the KL divergence is likely to overestimate the tails of the density \citep{Shang13}. To address this issue, \cite{ZK11} and \cite{Shang16} suggested that localized bandwidths should improve the estimation accuracy of error density for the symmetrical unimodal distribution. The kernel density estimator with localized bandwidths assigns small bandwidths to the observations in the high-density region and large bandwidths to the observations in the low-density region. The localized error-density estimator can be written as
\begin{equation*}
\widehat{f}\left(\varepsilon_i;\tau,\tau_{\varepsilon}\right) = \frac{1}{n-1}\sum^n_{\substack{j=1\\ j\neq i}}\frac{1}{\tau\left(1+\tau_{\varepsilon}|\widehat{\varepsilon}_j|\right)}\phi\left(\frac{\widehat{\varepsilon}_i-\widehat{\varepsilon}_j}{\tau\left(1+\tau_{\varepsilon}|\widehat{\varepsilon}_j|\right)}\right),
\end{equation*}
where $\tau(1+\tau_{\varepsilon}|\widehat{\varepsilon}_j|)$ is the bandwidth assigned to $\widehat{\varepsilon}_j$, for $j=1,2,\dots,n$ and the vector of parameters is now $(h, \tau, \tau_{\varepsilon})$. The adaptive random-walk Metropolis algorithm, described in Section~\ref{sec:3.3}, can be used for sampling parameters, where the prior density of $\tau_{\varepsilon}\sim U(0,1)$.

\section{Simulation}\label{sec:4}

The principal aim of this section is to illustrate the proposed method through simulated data. We consider two cases in which simulated curves are smooth in Section~\ref{sec:smooth} and in which simulated curves are rough in Section~\ref{sec:rough}. Using the error criteria in Sections~\ref{sec:4.1} and~\ref{sec:4.2}, we assess the estimation accuracies of the regression function and error density, respectively. In Section~\ref{sec:4.4}, we conduct a prior sensitivity analysis.

\subsection{Criteria for assessing estimation and prediction accuracies of the regression function}\label{sec:4.1}

To measure the estimation accuracy of the regression function, we first calculate the averaged mean squared error (AMSE) between the true regression function $g(\cdot)$ and the estimated regression function $\widehat{g}(\cdot)$. This is expressed as
\begin{align*}
\text{AMSE} = \text{E}\left\{\left[g(\mathcal{X}; \mathcal{Z}) - \widehat{g}(\mathcal{X}; \mathcal{Z})\right]^2\right\}&=\text{E}\left(\text{E}\left\{\left[g\left(\mathcal{X}; \mathcal{Z}\right)-\widehat{g}\left(\mathcal{X}; \mathcal{Z}\right)\right]^2\big|\mathcal{X}, \mathcal{Z}\right\}\right) \\
&\approx \frac{1}{n}\sum^n_{i=1}\text{E}\left\{\left[g(\mathcal{X}; \mathcal{Z})-\widehat{g}(\mathcal{X}; \mathcal{Z})\right]^2\big|\mathcal{X}=\mathcal{X}_i, \mathcal{Z} = \mathcal{Z}_i\right\}\\
&\approx\frac{1}{n}\frac{1}{B}\sum^n_{i=1}\sum^B_{s=1}\left[g(\mathcal{X}_i; \mathcal{Z}_i) - \widehat{g}^s(\mathcal{X}_i; \mathcal{Z}_i)\right]^2,
\end{align*}
where $B=100$ represents the number of replications.

To measure the prediction accuracy of the regression function, we calculate the averaged mean squared prediction error (AMSPE) between the holdout regression function $g(\mathcal{X}_{\text{new}}; \mathcal{Z}_{\text{new}})$ and the predicted regression function $\widehat{g}(\mathcal{X}_{\text{new}}; \mathcal{Z}_{\text{new}})$. This is expressed as
\begin{align*}
\text{AMSPE} &= \text{E}\left\{\left[g(\mathcal{X}_{\text{new}}; \mathcal{Z}_{\text{new}}) - \widehat{g}(\mathcal{X}_{\text{new}}; \mathcal{Z}_{\text{new}})\right]^2\right\}\\
&\approx \frac{1}{\eta}\frac{1}{B}\sum^{\eta}_{j=1}\sum^B_{s=1}\left[g(\mathcal{X}_j; \mathcal{Z}_j) - \widehat{g}^s(\mathcal{X}_j; \mathcal{Z}_j)\right]^2,
\end{align*}
where $\eta$ represents the length of prediction samples. 

\subsection{Criteria for assessing error-density estimation}\label{sec:4.2}

To measure the difference between the true error density $f(\varepsilon)$ and the estimated error density $\widehat{f}(\varepsilon)$, we consider two risk functions.  First, we calculate the mean integrated squared error (MISE), which is given by
\begin{eqnarray*}
\text{MISE}\left[\widehat{f}(\varepsilon)\right] = \int^b_{\varepsilon=a} \left[f(\varepsilon)-\widehat{f}(\varepsilon)\right]^2d\varepsilon,
\end{eqnarray*}
for $\varepsilon\in [a,b]$. For each replication, the MISE can be approximated at 1,001 grid points bounded between a closed interval, such as $[-10,10]$. These can be expressed as
\begin{eqnarray*}
\text{MISE}\left[\widehat{f}(\varepsilon)\right]\approx \frac{1}{50}\sum^{1001}_{i=1}\left\{f\Big[-10+\frac{(i-1)}{50}\Big]-\widehat{f}\Big[-10+\frac{(i-1)}{50}\Big] \right\}^2.
\end{eqnarray*}
From an average of 100 replications, the approximated mean integrated squared error (AMISE) is used to assess the estimation accuracy of error density. The AMISE is defined as
\begin{eqnarray*}
\text{AMISE}=\frac{1}{B}\sum^B_{b=1}\text{MISE}_b.
\end{eqnarray*}

The second risk function is the KL divergence, given by
\begin{eqnarray*}
\text{KL}\left(\widehat{f}, f\right) = -\text{E}\int_{\varepsilon=a}^b f(\varepsilon)\ln \widehat{f}(\varepsilon)d\varepsilon.
\end{eqnarray*}
With $f$ fixed in simulation, $\text{KL}\left(\widehat{f}, f\right)$ is minimized for $f=\widehat{f}$ \citep{Kullback59}.

\subsection{Simulation of smooth curves}\label{sec:smooth}

\begin{figure}[!ht]
\centering
\subfloat[Raw functional curves]
{\includegraphics[width=8.5cm]{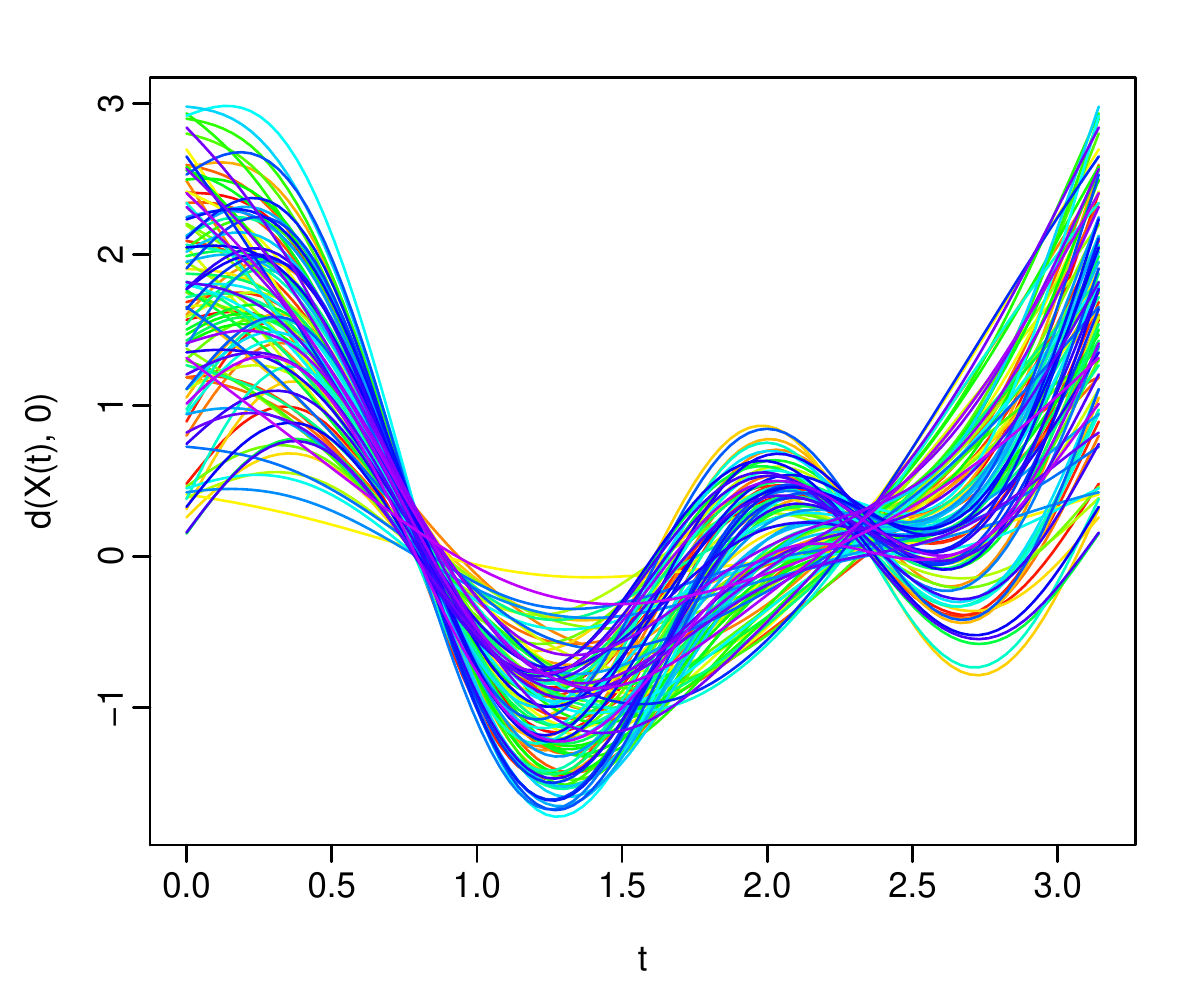}\label{fig:1a}}
\qquad
\subfloat[First-order derivative of the curves]
{\includegraphics[width=8.5cm]{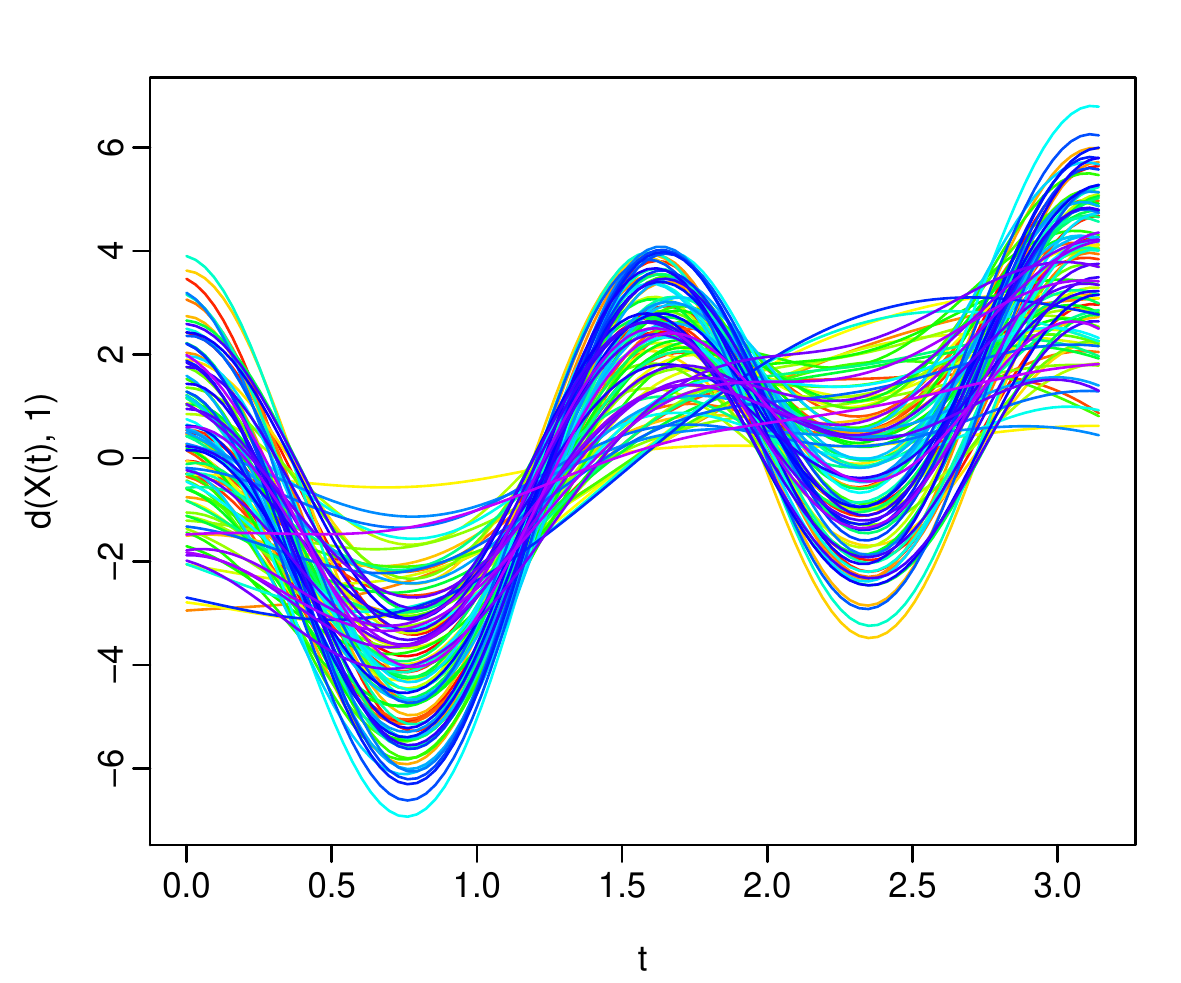}\label{fig:1b}}
\caption{100 simulated smooth curves.}\label{fig:1}
\end{figure}

We describe the construction of the simulated data. First, we build simulated discretized curves as
\begin{equation}
\mathcal{X}_i(t_j) = a_i\cos(2t_j) + b_i\sin(4t_j) + c_i\Big(t_j^2-\pi t_j + \frac{2}{9}\pi^2\Big), \qquad i=1,2,\dots,n,\label{eq:simu}
\end{equation}
where $t$ represents the function support range and $0\leq t_1\leq t_2\leq \dots \leq t_{100}\leq \pi$ are equispaced points within the function support range, $a_i$, $b_i$, $c_i$ are independently drawn from a uniform distribution on $[0,1]$, and $n$ represents the sample size. The functional form of~\eqref{eq:simu} is taken from \cite{FVV10} and \cite{Shang13, Shang13b}. Figure~\ref{fig:1} presents a replication of 100 simulated smooth curves, along with the first-order derivative of the curves approximated by B-splines. 

\newpage
Once the curves are formed, we define $\mathcal{Z}$ as the first-order derivative of the curves \citep[see also][]{MP09}. We compute the response variable through the following steps:
\begin{enumerate}
\item[1)] Construct a regression-function operator $g$, which performs the mapping from function-valued space to real-valued space. The regression-function operator $g$ is expressed as
\begin{equation*}
g(\mathcal{X}_i; \mathcal{Z}_i) = 10\times \left(a_i^2-b_i^2\right).
\end{equation*}
\item[2)] Generate $\varepsilon_1,\varepsilon_2,\dots,\varepsilon_n$, which are independently drawn from mixture normal distributions. To highlight the possible non-normality of the error density, we consider three different error densities previously introduced by \cite{MW92}:
\begin{enumerate}
\item unimodal symmetric distribution, such as $t_5$;
\item skewunimodal distribution: $0.2\times \text{N}(0,1) + 0.2\times \text{N}\left(0.5,(\frac{2}{3})^2\right) + 0.6\times \text{N}\left(13/12,(\frac{5}{9})^2\right)$;
\item skewbimodal distribution: $0.75\times \text{N}(0,1) + 0.25\times \text{N}\left(1.5,(\frac{1}{3})^2\right)$.
\end{enumerate}
\item[3)] Compute the corresponding responses: $y_i = g(\mathcal{X}_i; \mathcal{Z}_i) + \varepsilon_i$, for $i=1,2,\dots,n$.
\end{enumerate}

\subsubsection{Estimating the regression function}

For a given data triplet ($\mathcal{X}, \mathcal{Z}, y$) and a bandwidth $h$, we compute the discrepancy between $g(\cdot)$ and $\widehat{g}(\cdot)$. We use the following Monte-Carlo scheme:
\begin{itemize}
\item Build 100 replications $\left[(\mathcal{X}_i^{s}, \mathcal{Z}_i^{s}, y_i^{s})_{i=1,\dots,n}\right]_{s=1,\dots,100}$;
\item Compute 100 estimates $\left[g(\cdot)-\widehat{g}^s(\cdot)\right]_{s=1,\dots,100}$, where $\widehat{g}^s(\cdot)$ is the functional NW estimator of the regression function computed over the $s$\textsuperscript{th} replication;
\item Obtain the AMSE and AMSPE by averaging over 100 replications of MSE and MSPE.
\end{itemize}

Table~\ref{tab:1} presents the AMSEs and AMSPEs for the estimated conditional mean, where the error density is estimated by a kernel error density with a global bandwidth and localized bandwidths. The functional partial linear model considers two function-valued random variables, $\mathcal{X}$ and $\mathcal{Z}$. In contrast, functional principal component regression and functional nonparametric regression models consider only the original functional curves, $\mathcal{X}$. Thus, the functional partial linear model produces improved estimation and forecast accuracies than functional principal component regression and functional nonparametric regression.  It is also advantageous to estimate error density with localized bandwidths to achieve the best estimation and forecast accuracies of the regression function. An exception is in the skewbimodal error density, where the difference between the two bandwidth procedures is quite small. Among the three types of semi-metrics, the AMSEs are smaller for the two semi-metrics based on derivatives, compared with the semi-metric based on FPCA. Following the early work by \cite{Shang13b}, we use three retained principal components that seem to be sufficient to capture the primary mode of variation in this example. The semi-metric based on the second derivative has the smallest AMSE and AMSPE among the three semi-metrics.

\begin{center}
\tabcolsep 0.05in
\begin{longtable}{@{}ll|lll|lll|lll@{}}
\caption{Comparisons of the estimation and forecast accuracies of the regression function for the functional partial linear model, functional principal component regression, and functional nonparametric regression for the three choices of semi-metric. The underscore G and L represent a global bandwidth or localized bandwidths for estimating error density. The smallest errors are highlighted in bold.}\label{tab:1} \\
\toprule
& & \multicolumn{9}{c}{Type of semi-metric} \\
& & \multicolumn{3}{c}{$1^{\text{st}}$ derivative} & \multicolumn{3}{c}{$2^{\text{nd}}$ derivative} &\multicolumn{3}{c}{FPCA$_{Z=3}$} \\ 
& FPCR & FPLM$_{\text{G}}$ & FPLM$_{\text{L}}$ & FNP & FPLM$_{\text{G}}$ & FPLM$_{\text{L}}$ & FNP & FPLM$_{\text{G}}$ & FPLM$_{\text{L}}$ & FNP \\\toprule
\endfirsthead
\toprule
& & \multicolumn{9}{c}{Type of semi-metric} \\
& & \multicolumn{3}{c}{$1^{\text{st}}$ derivative} & \multicolumn{3}{c}{$2^{\text{nd}}$ derivative} &\multicolumn{3}{c}{FPCA$_{Z=3}$} \\ 
& FPCR & FPLM$_{\text{G}}$ & FPLM$_{\text{L}}$ & FNP & FPLM$_{\text{G}}$ & FPLM$_{\text{L}}$ & FNP & FPLM$_{\text{G}}$ & FPLM$_{\text{L}}$ & FNP \\\toprule
\endhead
\hline \multicolumn{11}{r}{{Continued on next page}} \\ 
\endfoot
\endlastfoot
\hspace{.02in}{\underline{$t_5$}} \\
AMSE & 1.148 & 1.039 & 0.993 & 5.382 & 1.032 & \textBF{0.969} & 5.039 & 1.164 & 1.097 & 5.713 \\
AMSPE & 1.332 & 1.361 & 1.285 & 5.247 & 1.329 & \textBF{1.235} & 5.055 &1.541 & 1.452 & 5.484 \\
\\
\hspace{.02in}{\underline{Skewunimodal}} \\
AMSE & 1.600 & 1.237 & 1.216 & 5.847 & 1.148 & \textBF{1.136} & 5.353 & 1.337 & 1.320 & 6.182 \\
AMSPE &1.783 & 1.366 & 1.371 & 5.746 & 1.313 & \textBF{1.297} & 5.383 & 1.551 & 1.519 & 5.980 \\
\\
\hspace{.02in}{\underline{Skewbimodal}} \\
AMSE & 1.235 & 1.010 & 1.006 & 5.450 & \textBF{0.971} & 0.973 & 5.017 & 1.130 & 1.116 & 5.771 \\
AMSPE & 1.416 & 1.213 & 1.199 & 5.334 & \textBF{1.144} & 1.145 & 5.041 & 1.363 & 1.353 & 5.569 \\
\bottomrule
\end{longtable}
\end{center}

\vspace{-.2in}

\begin{figure}[!ht]
\centering
\subfloat[A global bandwidth]
{\includegraphics[width=7.9cm]{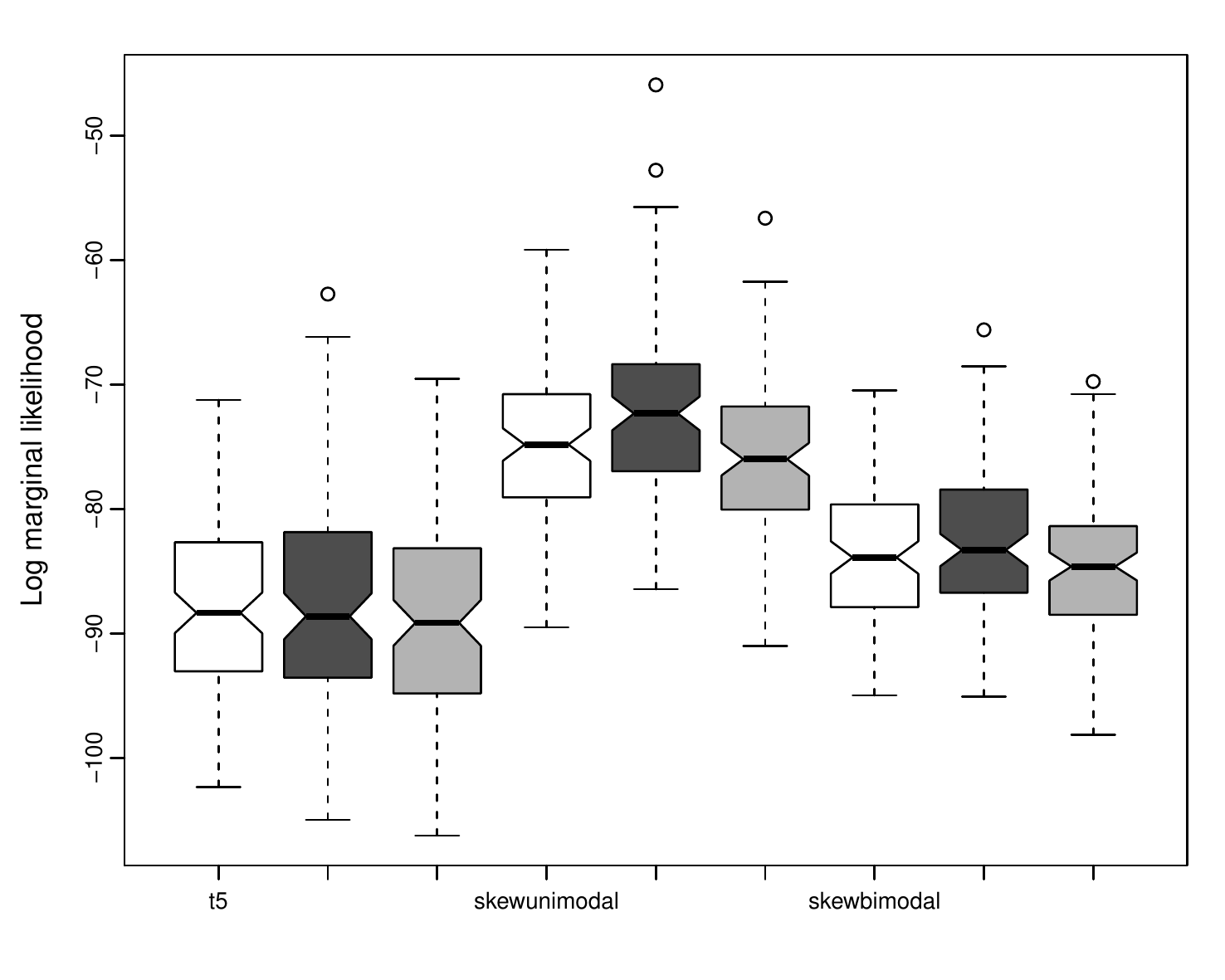}\label{fig:2a}}
\qquad
\subfloat[Localized bandwidths]
{\includegraphics[width=7.9cm]{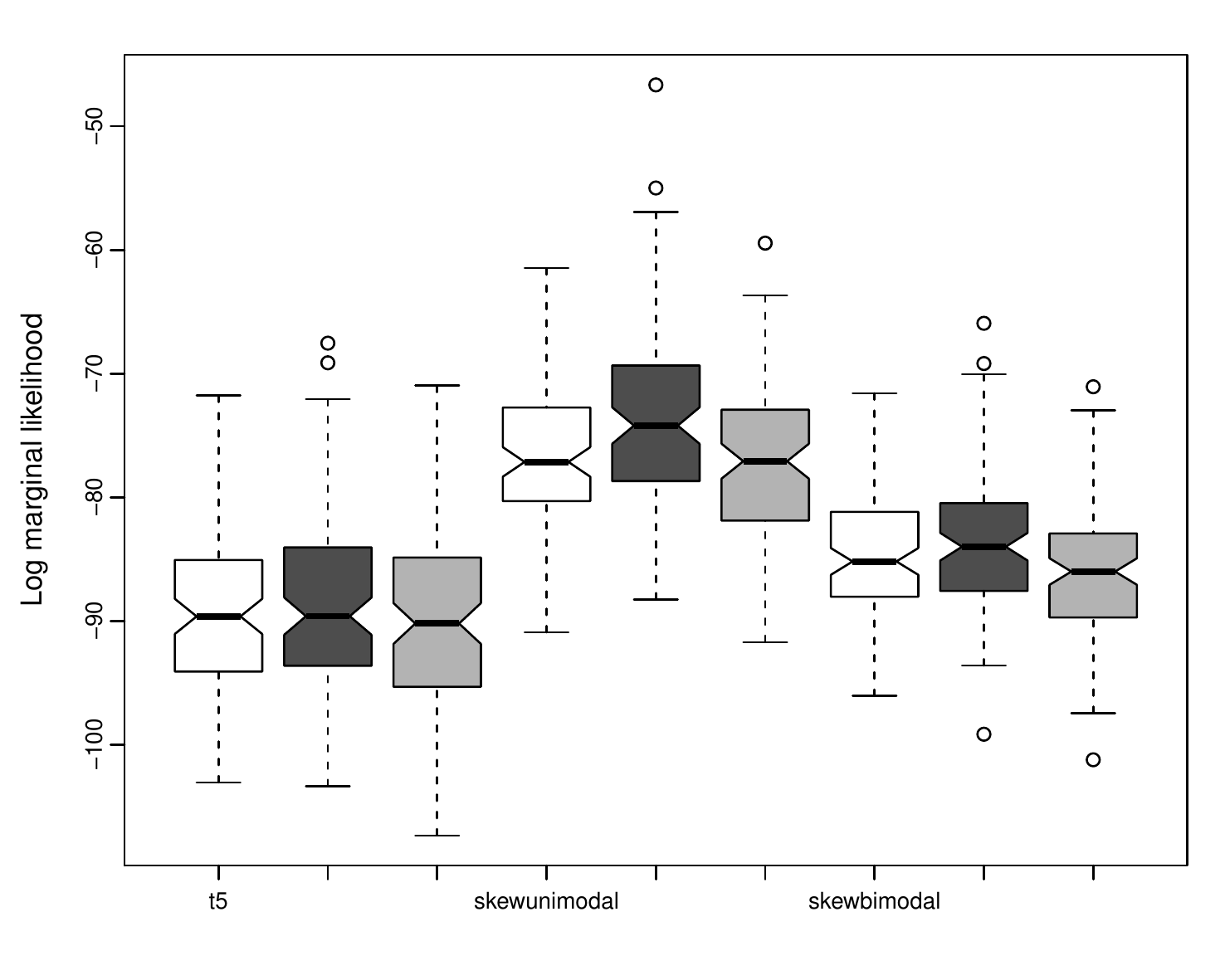}\label{fig:2b}}
\caption{Comparisons of the LMLs in the functional partial linear model with a global bandwidth and localized bandwidths for the three choices of semi-metric. The semi-metric based on the first derivative is shown in white, the semi-metric based on the second derivative is shown in dark grey, and the semi-metric based on the FPCA is shown in light grey.}\label{fig:2}
\end{figure}

Figure~\ref{fig:2} presents the log marginal likelihoods (LMLs) for the semi-metrics in all three error densities considered. Again, the semi-metric based on the second derivative has the largest LMLs and is thus considered the optimal semi-metric in this example.

\subsubsection{Estimating the error density}

With a set of residuals and a fixed residual bandwidth, it is possible to apply a univariate kernel-density estimator and compute the discrepancy between $f(\varepsilon)$ and $\widehat{f}(\varepsilon)$ by using the following Monte-Carlo scheme:
\begin{itemize}
\item Compute 100 replications of residuals $\left[y_i^s-\widehat{g}_i^s(\cdot)\right]_{s=1,\dots,100}$;
\item Apply a univariate kernel density to estimate error density, where the residual bandwidth is estimated by the Bayesian method for 100 replications;
\item Compute the MISE and KL divergence between the true error density $f^s(\varepsilon)$ and estimated error density $\widehat{f}^s(\varepsilon)$ for $s=1,2,\dots,100$;
\item Obtain the AMISE and averaged KL divergence by averaging over 100 replications of MISE and KL divergence.
\end{itemize}

Table~\ref{tab:2} presents AMISEs and averaged KL divergences for the kernel error density with bandwidths estimated by Bayesian methods with a global bandwidth and localized bandwidths. For the three error densities, the kernel error density with localized bandwidths produces smaller AMISEs and smaller averaged KL divergences than the kernel error density with a global bandwidth.

\begin{table}[!htbp]
\centering
\tabcolsep 0.12cm
\caption{AMISEs and averaged KL divergences for the error-density estimation in the functional partial linear model with a global bandwidth and localized bandwidths for the three choices of semi-metric. The underscore G and L represent a global bandwidth and localized bandwidths for estimating error density. The smallest errors are highlighted in bold.}\label{tab:2}
\begin{tabular}{@{}llllllll@{}}\toprule
& \multicolumn{3}{c}{Type of semi-metrics (G)} & & \multicolumn{3}{c}{Type of semi-metrics (L)} \\
$f(\varepsilon)$ & $1^{\text{st}}$ derivative & $2^{\text{nd}}$ derivative & FPCA$_{Z=3}$ & & $1^{\text{st}}$ derivative & $2^{\text{nd}}$ derivative & FPCA$_{Z=3}$ \\\midrule
\underline{AMISE} \\
$t_5$ & 0.0015 & 0.0015 & 0.0019 & & \textBF{0.0008} & \textBF{0.0008} & 0.0008 \\
skewunimodal & 0.0088 & 0.0084 & 0.0088 & & 0.0074 & \textBF{0.0073} & 0.0075 \\
skewbimodal & 0.0026 & 0.0024 & 0.0026 & & 0.0024 & \textBF{0.0022} & 0.0026 \\
\\
\multicolumn{4}{l}{\hspace{-.13in} {\underline{Averaged KL divergence}}} \\
$t_5$ & 0.1427 & 0.1421 & 0.1957 & & 0.1299 & \textBF{0.1272} & 0.1391 \\
skewunimodal & 0.6205 & 0.5486 & 0.6737 & & 0.4996 & \textBF{0.4636} & 0.5695 \\
skewbimodal & 0.4247 & 0.3905 & 0.4736 & & 0.2990 & \textBF{0.2756} & 0.4112 \\
\bottomrule
\end{tabular}
\end{table}

\subsection{Diagnostic check}\label{sec:4.4}

\begin{figure}[!htbp]
\centering
{\includegraphics[width=8.21cm]{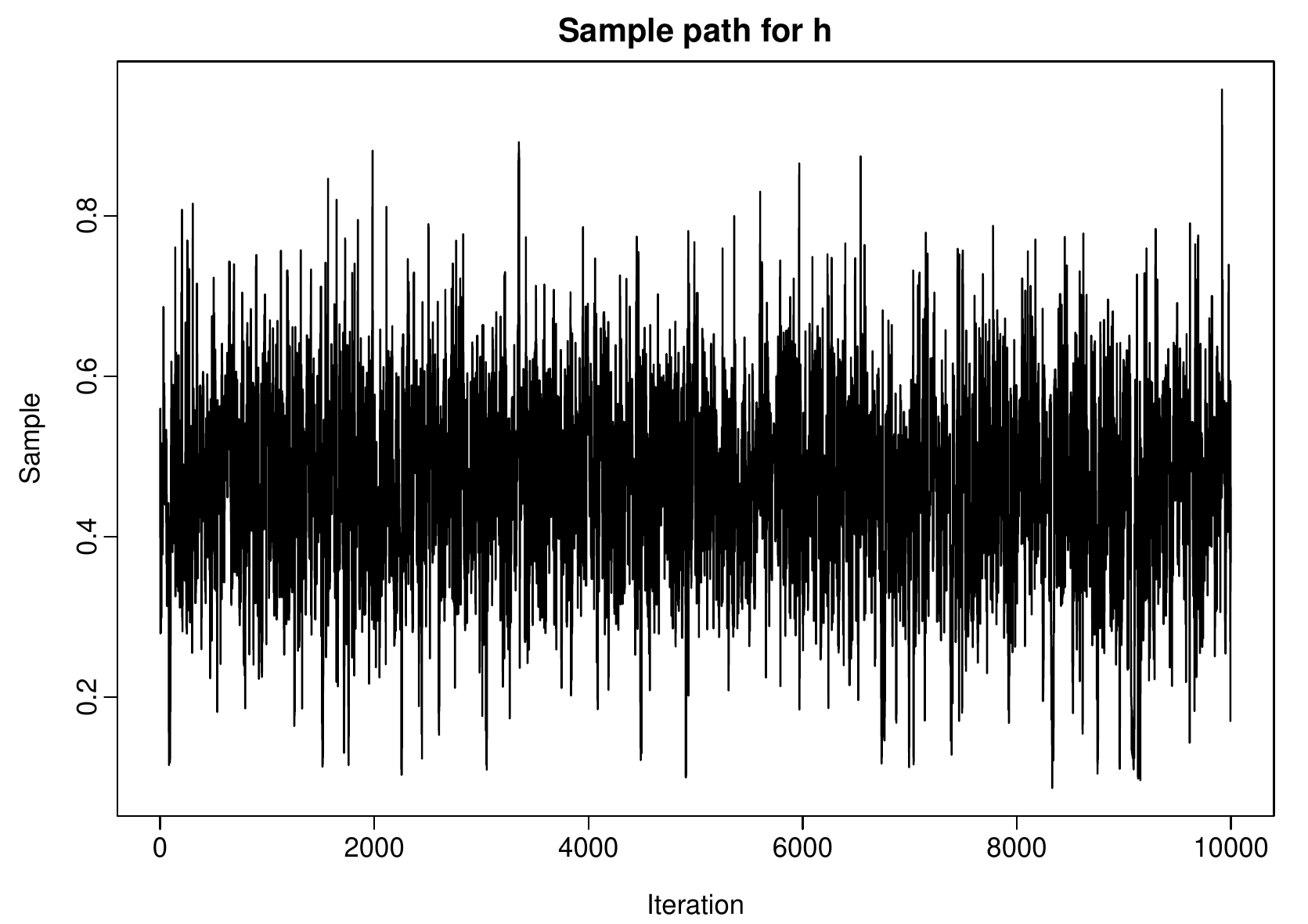}}
\quad
{\includegraphics[width=8.21cm]{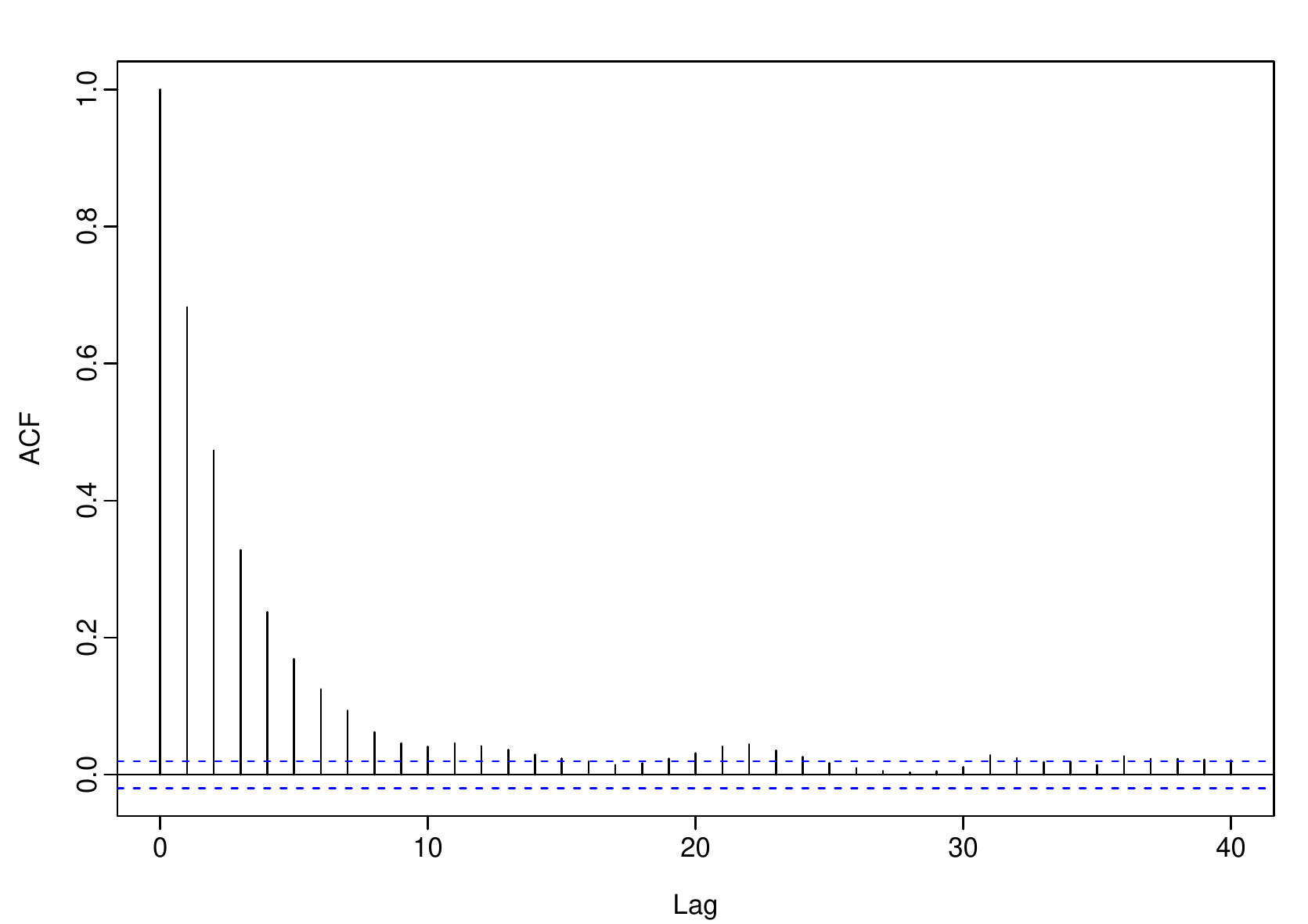}}
\\
{\includegraphics[width=8.21cm]{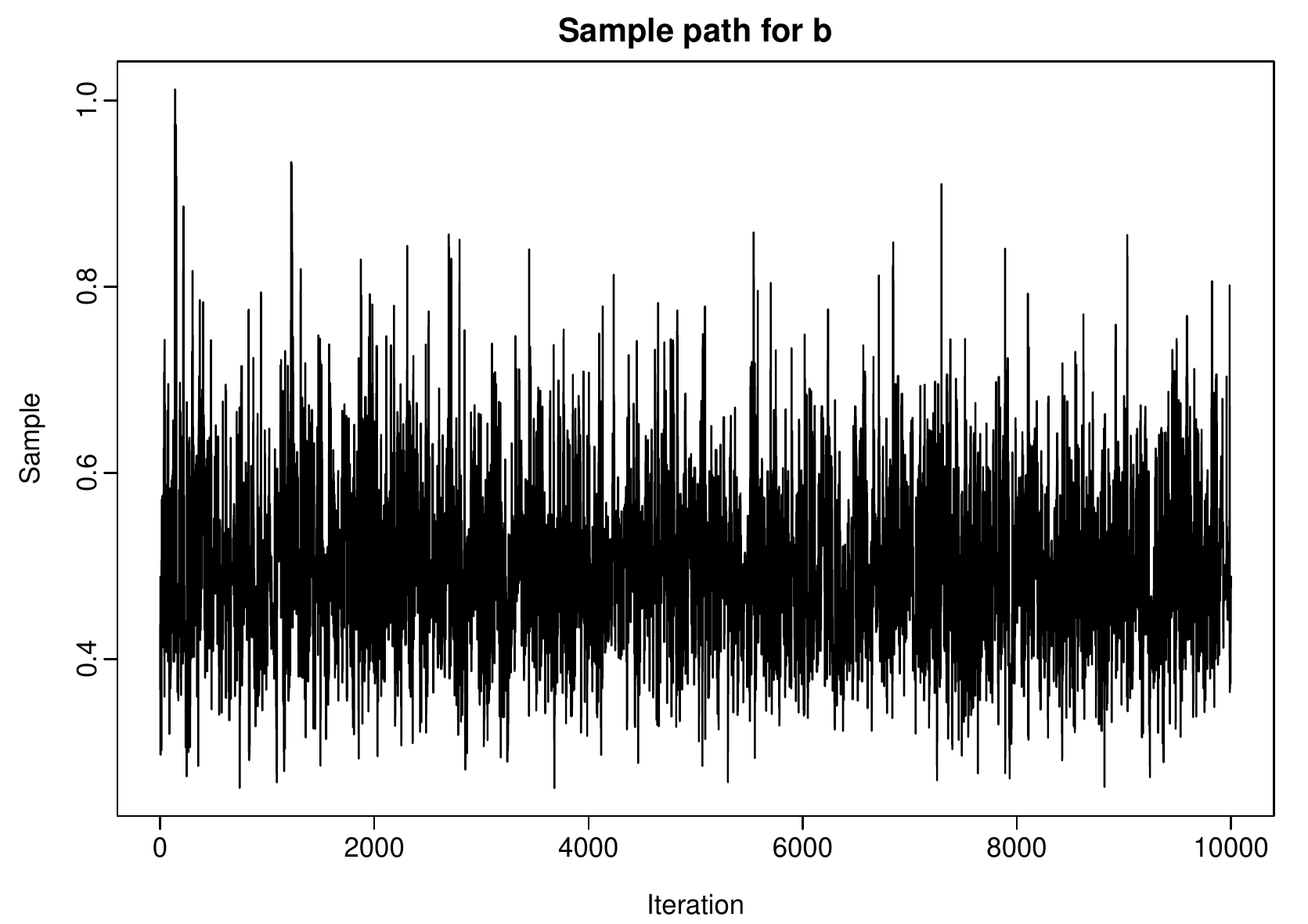}}
\quad
{\includegraphics[width=8.21cm]{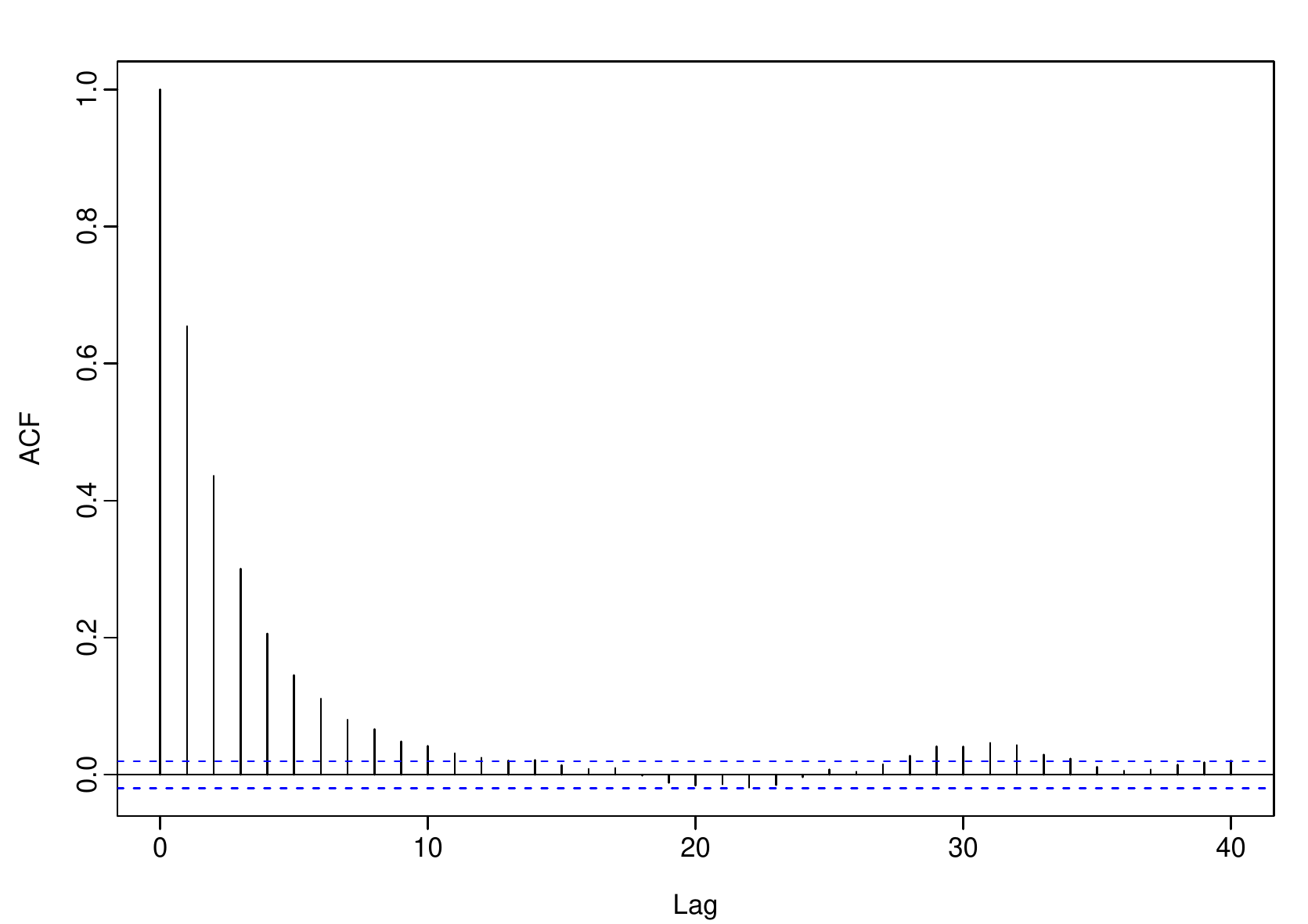}}
\caption{MCMC sample paths and ACFs of the sample paths with $t_5$ error density.}\label{fig:diag}
\end{figure}

As a demonstration with one replication, we plot the MCMC sample paths of the parameters on the left panel of Figure~\ref{fig:diag} and the autocorrelation functions (ACFs) of these sample paths on the right panel of Figure~\ref{fig:diag}. With the $t_5$ error density, these plots demonstrate that the sample paths are mixed reasonably well. Table~\ref{tab:diag} summarizes the ergodic averages, 95\% Bayesian credible intervals (CIs), standard error (SE), batch mean SE, and simulation inefficiency factor (SIF) values. Considered previously by \cite{KSC98} and \cite{MY00}, the SIF can be interpreted as the number of draws needed to have iid observations. Based on the SIF values, we find that the simulation chain converges very well. 

\begin{table}[!bhtp]
\tabcolsep 0.27in
\centering
\caption{MCMC results of the bandwidth estimation under the prior density of IG$(\alpha=1,\beta=0.05)$ with $t_5$ error density.}\label{tab:diag}
\begin{tabular}{@{}llllll@{}}\toprule
\multicolumn{3}{l}{\hspace{-.255in}{Prior density: IG$(\alpha=1, \beta=0.05)$}} & SE & Batch-mean SE & SIF \\
Parameter & Mean & 95\% Bayesian CIs & & & \\\midrule
$h$ & 0.4656 & (0.2136, 0.7026) & 0.1217 & 0.3175 & 6.80 \\
$b$ & 0.5000 & (0.3377, 0.7078) & 0.0967 & 0.2300 & 5.65 \\\bottomrule
\end{tabular}
\end{table}

\subsection{Simulation of rough curves}\label{sec:rough}

\begin{figure}[!htbp]
\centering
\subfloat[Raw functional curves]
{\includegraphics[width=8.5cm]{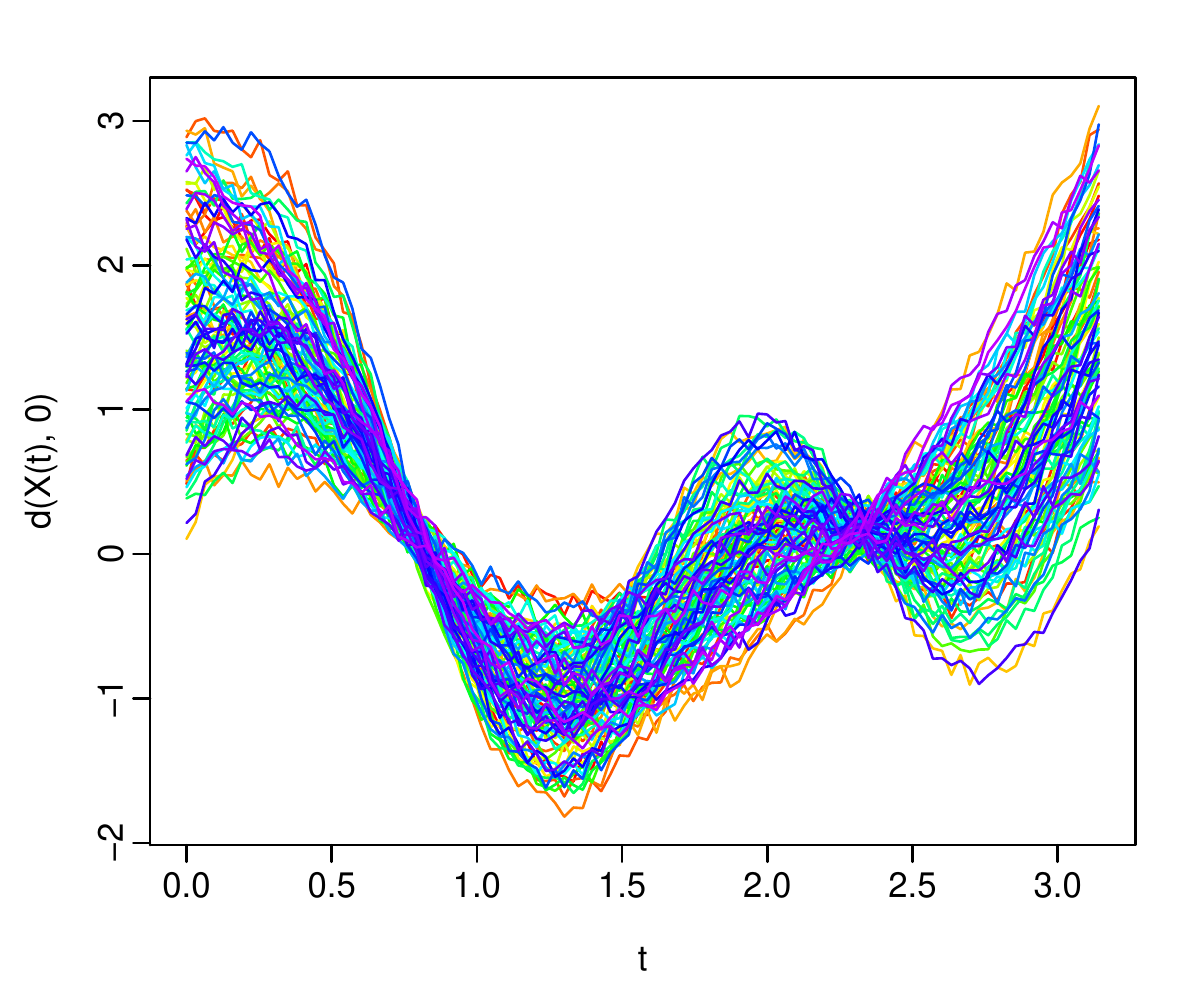}}
\qquad
\subfloat[The first-order derivative of the curves]
{\includegraphics[width=8.5cm]{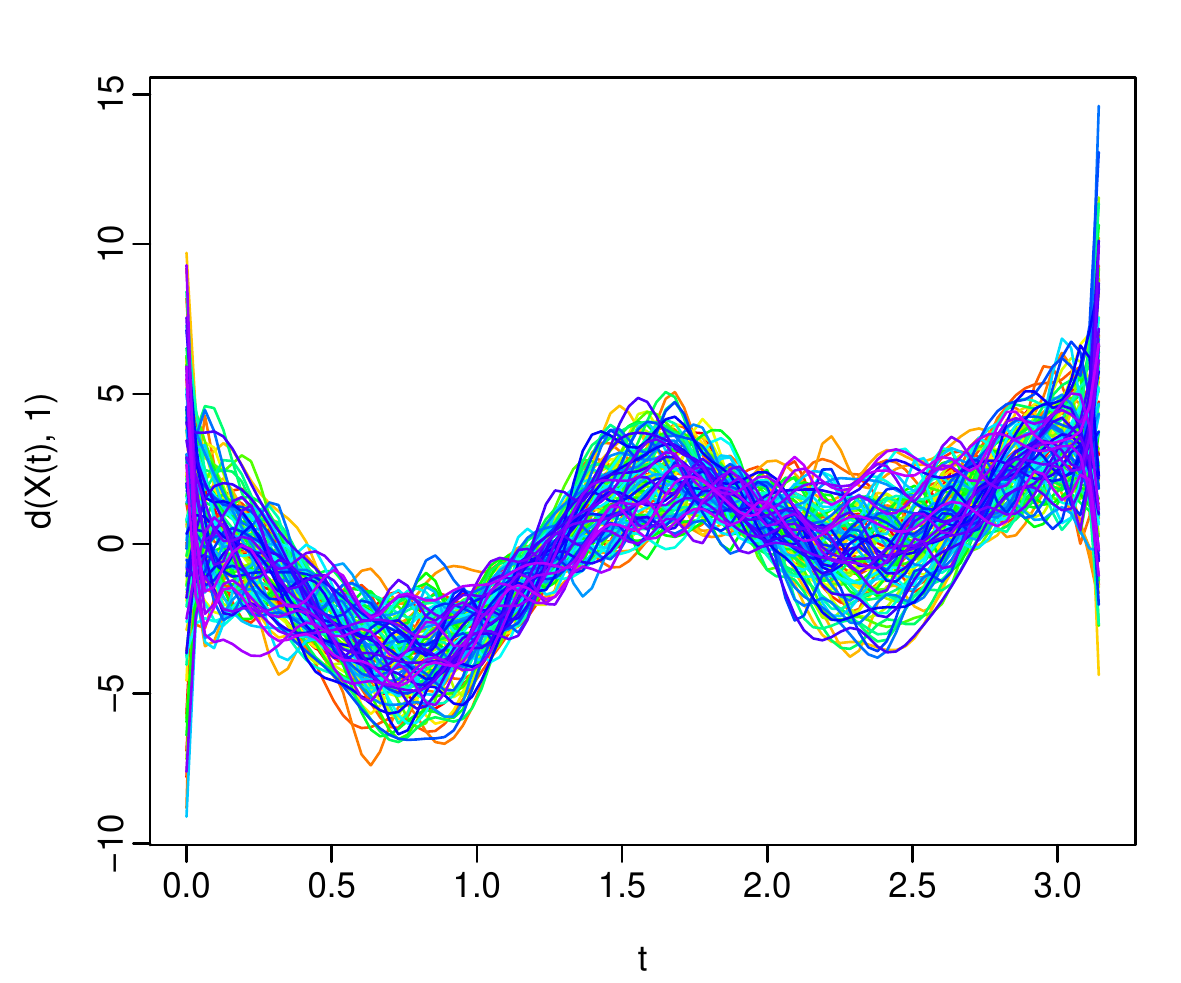}}
\caption{100 simulated rough curves.}\label{fig:rough}
\end{figure}

In this simulation study, we consider the same functional form as given in~\eqref{eq:simu}, but add one extra variable $d_j\sim U(-0.1, 0.1)$ in the construction of curves. This functional form is taken from \citet[][Section 4.2]{Shang13b}. Figure~\ref{fig:rough} presents the simulated curves for one replication, along with the first-order derivative of the curves. 

\begin{figure}[!ht]
\centering
\includegraphics[width=13cm]{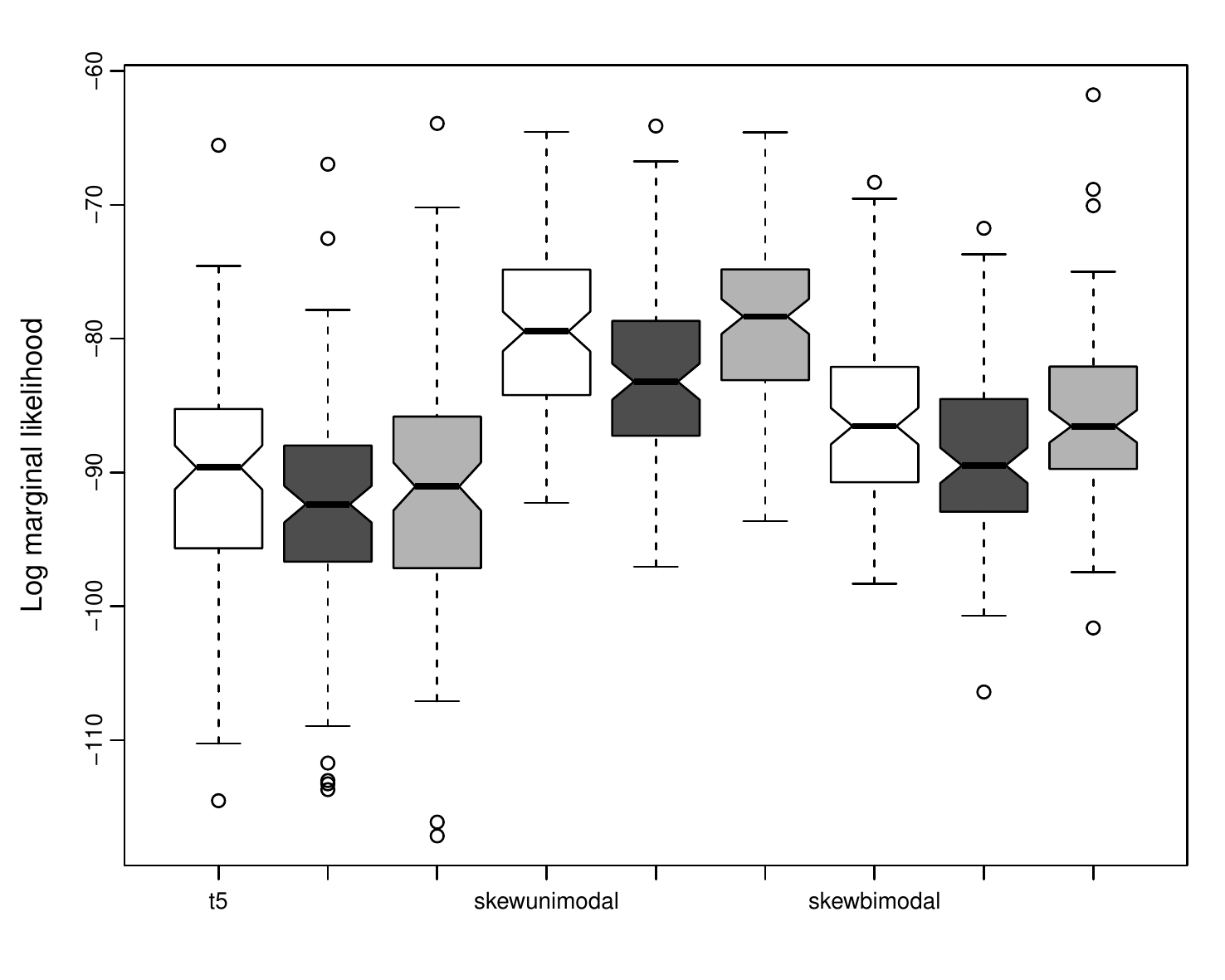}
\caption{Comparisons of the LMLs in the functional partial linear model with a global bandwidth for the three choices of semi-metric. The semi-metric based on the first derivative is shown in white, the semi-metric based on the second derivative is shown in dark grey, and the semi-metric based on the FPCA is shown in light grey.}\label{fig:Chib_rough}
\end{figure}

From the smooth curves presented in Figure~\ref{fig:2}, we found the difference in LML between a global bandwidth and localized bandwidths to be quite small for the functional partial linear model. Therefore, in Figure~\ref{fig:Chib_rough}, we report only the LMLs for the functional partial linear model with a global bandwidth. The LMLs are computed for the three semi-metrics in all three error densities considered,  and the overall optimal semi-metric is determined based on 100 replications. With the median LML, the semi-metric based on the first derivative has the largest LML for the $t_5$ error density, followed closely by the semi-metric based on the FPCA with three retained components. In the skewunimodal and skewbimodal error densities, the semi-metric based on the FPCA gives the largest LML and is thus considered the optimal semi-metric in these two error densities.

\section{Spectrometric data analysis}\label{sec:5}

\begin{figure}[!ht]
\centering
\includegraphics[width=12.3cm]{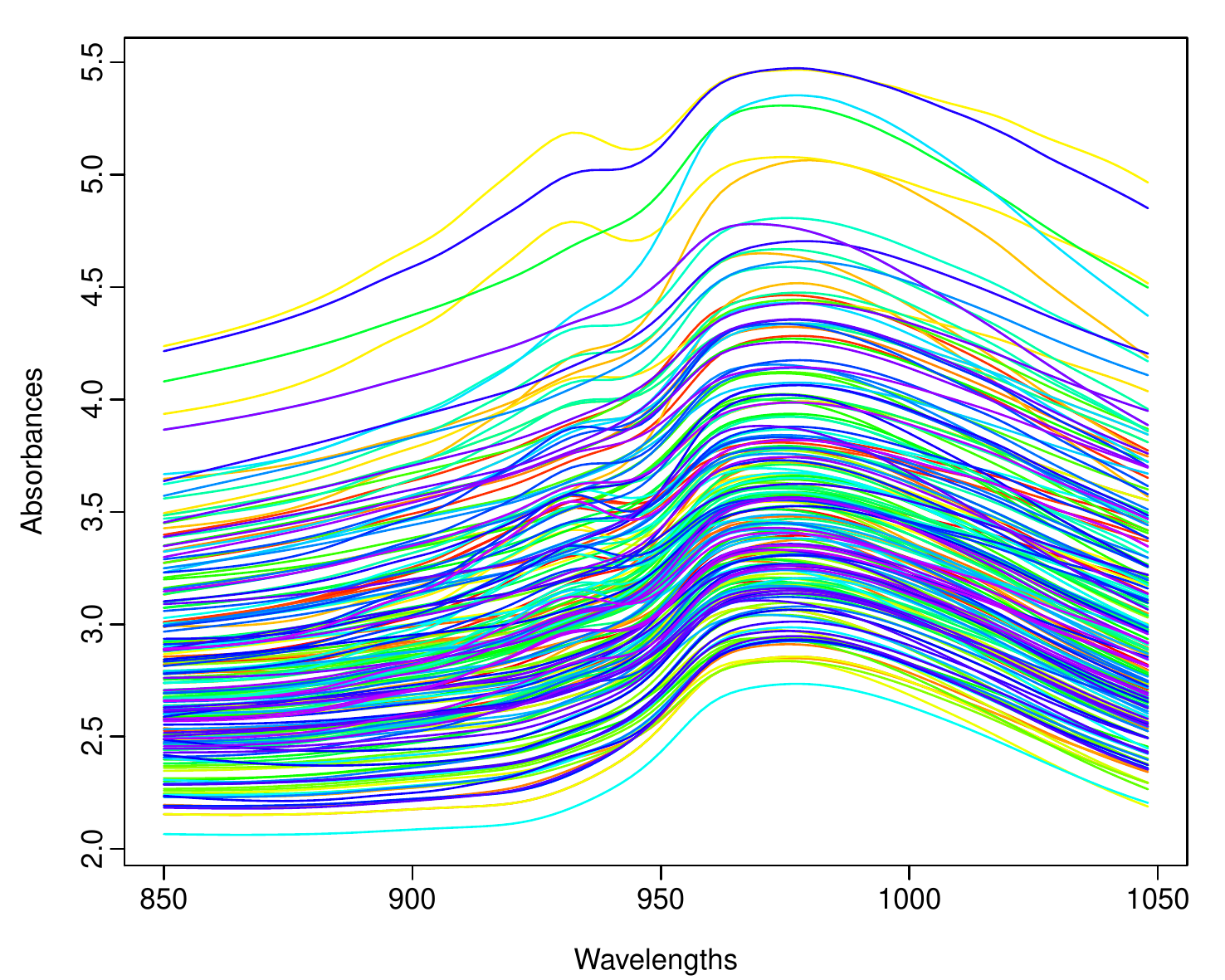}
\caption{Graphical display of spectrometric curves.}\label{fig:3}
\end{figure}

This dataset focuses on the prediction of the fat content in meat samples based on near-infrared (NIR) absorbance spectra. The dataset was obtained from \url{http://lib.stat.cmu.edu/datasets/tecator} and has been studied by many researchers \citep[e.g.,][]{FV06, AV06}. Each food sample contains finely chopped pure meat with a percentage of fat content. For each unit $i$ (among 215 pieces of finely chopped meat), we observe one spectrometric curve, denoted by $\mathcal{X}_i$. The spectrometric curve measures the absorbance at a grid of 100 wavelengths, denoted by $\mathcal{X}_i=[\mathcal{X}_i(t_1),\dots,\mathcal{X}_i(t_{100})]$. For each unit $i$, we consider the first-order derivative of $\mathcal{X}_i$ and observe the fat content $y\in R$ by analytical chemical processing. Given a new spectrometric curve $(\mathcal{X}_{\text{new}}, \mathcal{X}^{'}_{\text{new}})$, we aim to predict the corresponding fat content $y_{\text{new}}$. A graphical display of the spectrometric curves is presented in Figure~\ref{fig:3}. 

To assess the point forecast accuracy of the functional partial linear model, we split the original samples into two subsamples \citep[see also][p.105]{FV06}. The first subsample is named the `learning sample', which contains the first 160 units $\{\mathcal{X}_i, \mathcal{Z}_i, y_i\}_{i=1,2,\dots,160}$. The second subsample is named the `testing sample', which contains the last 55 units $\{\mathcal{X}_l, \mathcal{Z}_l, y_l\}_{l=161,\dots,215}$. The learning sample allows us to build the regression model with the optimal smoothing parameter. The testing sample allows us to evaluate the prediction accuracy.

To measure estimation and prediction accuracies, we consider the root mean squared error (RMSE) and root mean squared prediction error (RMSPE). The errors are expressed as 
\begin{align*}
\text{RMSE}&=\sqrt{\frac{1}{160}\sum_{\omega=1}^{160}\left(y_{\omega}-\widehat{y}_{\omega}\right)^2}, \\ 
\text{RMSPE}&=\sqrt{\frac{1}{55}\sum_{\delta=161}^{215}\left(y_{\delta}-\widehat{y}_{\delta}\right)^2}. 
\end{align*}
For comparison, we also consider the functional principal component regression examined by \cite{RO07} and the functional nonparametric regression examined by \cite{FV06}.

As presented in Table~\ref{tab:3}, the functional partial linear model produces the smallest RMSE and RMSPE, in particular when the semi-metric is chosen based on the second derivative. The semi-metric based on the second derivative not only has the smallest RMSE and RMSPE, but it also has the largest LML. 

\begin{table}[!ht]
\centering
\tabcolsep 0.6cm
\begin{small}
\caption{Estimation and forecast accuracies for the functional partial linear model, functional principal component regression, and functional nonparametric regression. The smallest error measures and largest LML are shown in bold.}\label{tab:3}
\begin{tabular}{@{}lc|cc|cc|cc@{}}\toprule
& & \multicolumn{6}{|c}{Type of semi-metrics} \\\cmidrule{3-8}
& & \multicolumn{2}{c}{$1^{\text{st}}$ derivative} & \multicolumn{2}{c}{$2^{\text{nd}}$ derivative} & \multicolumn{2}{c}{FPCA$_{Z=3}$} \\ 
Criterion & FPCR  & FPLM & FNP & FPLM & FNP & FPLM & FNP \\\toprule
RMSE & 8.0832  &  2.4281  & 4.6014 & \textBF{1.5993} & 2.2899 & 7.3236 & 10.8779 \\
RMSPE & 9.1156 & 1.9191 & 4.7755 & \textBF{1.4075} & 1.9429 & 8.3378 & 11.2603 \\ 
LML        & & -360.21                &         &     \textBF{-291.37}   &          &  -524.48 &  \\
\bottomrule
\end{tabular}
\end{small}
\end{table}

\begin{figure}[!ht]
\centering
\includegraphics[width=\textwidth]{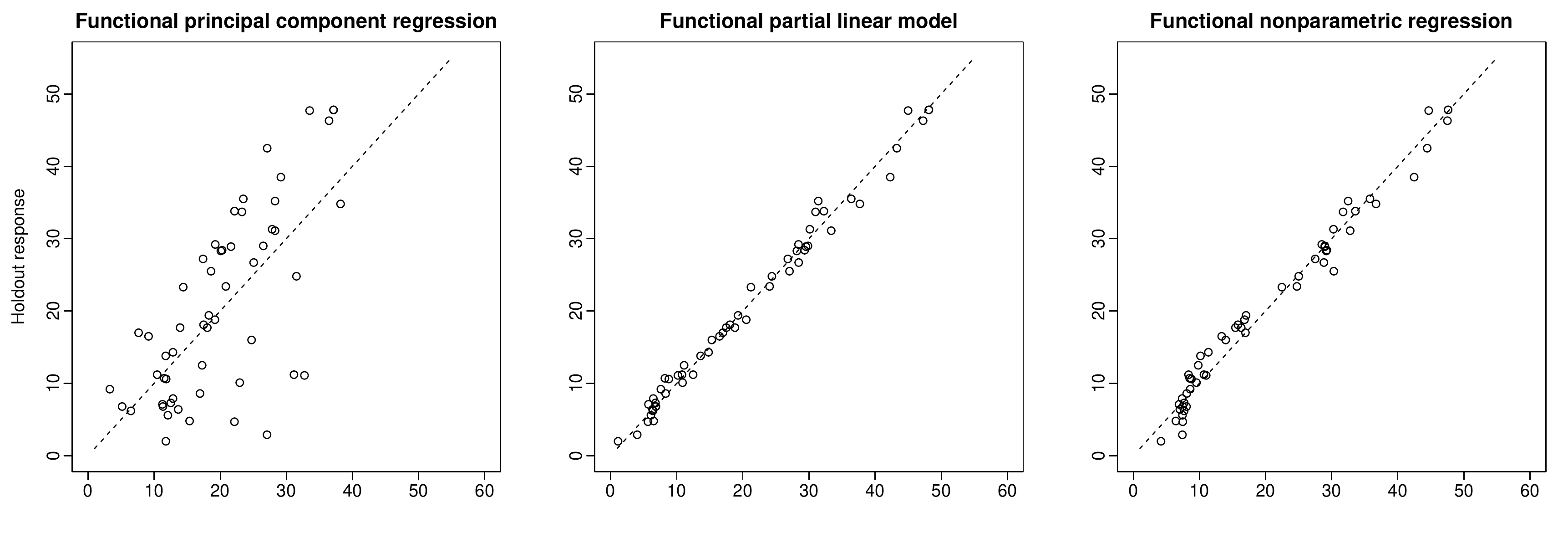}
\caption{Diagnostic check of the three functional regression models.}\label{fig:diag_2}
\end{figure}

To verify the model adequacy, we present several scatter plots of the holdout responses against the predicted responses in Figure~\ref{fig:diag_2}. Among the three regression models, the functional partial linear model provides the best prediction accuracy.

\begin{figure}[!ht]
\centering
\includegraphics[width=12cm]{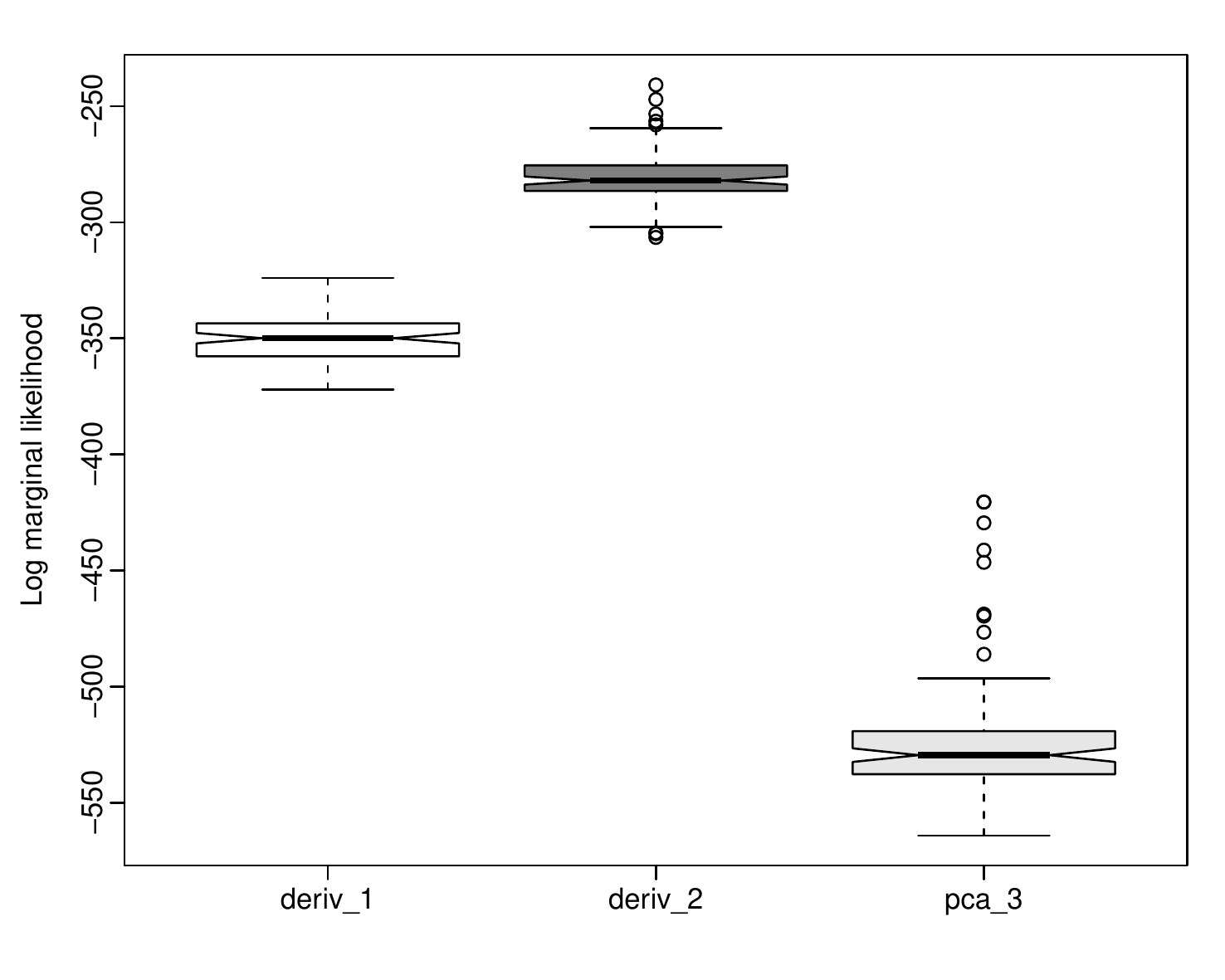}
\caption{Comparisons of the LMLs in the functional partial linear model with a global bandwidth for the three choices of semi-metric.}\label{fig:4}
\end{figure}

The original functional data being iid allows sampling with replacement to obtain a set of bootstrap samples. For each bootstrap sample, we use the first 160 pairs of data for estimation and the remaining 55 pairs of data for assessing the prediction accuracy. By repeating this procedure 100 times, we obtain the averaged RMSEs, and RMSPEs presented in Table~\ref{tab:4}. The functional partial linear model produces more accurate forecasts than functional principal component regression and functional nonparametric regression. In particular, the functional partial linear model produces the most accurate forecasts when the semi-metric is based on the second derivative. This optimal selection of semi-metric is further confirmed by the boxplot of LMLs presented in Figure~\ref{fig:4}. 

\begin{table}[!htb]
\centering
\tabcolsep 0.62cm
\caption{Estimation and forecast accuracy averaged over 100 replications for the functional partial linear model, functional principal component regression, and functional nonparametric regression. The smallest error measures are shown in bold.}\label{tab:4}
\begin{small}
\begin{tabular}{@{}lc|cccccc@{}}\toprule
& & \multicolumn{6}{c}{Type of semi-metrics} \\
& & \multicolumn{2}{c}{$1^{\text{st}}$ derivative} & \multicolumn{2}{c}{$2^{\text{nd}}$ derivative} & \multicolumn{2}{c}{FPCA$_{Z=3}$} \\ 
Criterion & FPCR & FPLM & FNP & FPLM & FNP & FPLM & FNP \\\toprule
RMSE  & 8.1011 &   2.2858  & 4.6449    & \textBF{1.5269} & 2.2343 & 7.3675 & 10.7080  \\
RMSPE  & 8.2234 &  2.3100 & 4.5655     & \textBF{1.6259} & 2.1947 & 7.8725 & 10.8047   \\
\bottomrule
\end{tabular}
\end{small}
\end{table}

There has been rapid development in functional regression models. Here, we consider a large set of functional regression models collected in \cite{GS14} and \cite{FV11}. These models include: 
\begin{inparaenum}
\item[(1)] restricted maximum likelihood-based functional linear model with a locally adaptive penalty \citep{CFS03}; 
\item[(2)] penalized functional regression \citep{GBC+11}; 
\item[(3)] functional principal component regression on the first few functional principal components \citep{RO07}; 
\item[(4)] linear model on the first $K$ functional principal components, where optimal $K$ is estimated by 20-fold bootstrap \citep{RS05}; 
\item[(5)] REML-based single-index signal regression with locally adaptive penalty \citep{Wood11}; 
\item[(6)] cross-validation based single-index signal regression \citep{ELM09}; 
\item[(7)] penalized partial least squares \citep{KBT08}; 
\item[(8)] least absolute shrinkage and selection operator penalized linear model on the first few functional principal components \citep{FHT10}; 
\item[(9)] functional nonparametric regression with NW estimator for estimating conditional mean \citep{FV06}; 
\item[(10)] functional nonparametric regression with NW estimator for estimating conditional median \citep{LLS11}; 
\item[(11)] functional nonparametric regression with NW estimator for estimating conditional mode \citep{FLV05}; 
\item[(12)] functional nonparametric regression with $k$ nearest neighbor estimator \citep{BFV09}; 
\item[(13)] functional nonparametric regression with most-predictive design points \citep{FHV10}; 
\item[(14)] functional partial linear model. 
\end{inparaenum}

\begin{figure}[!htbp]
\centering
\includegraphics[width=14cm]{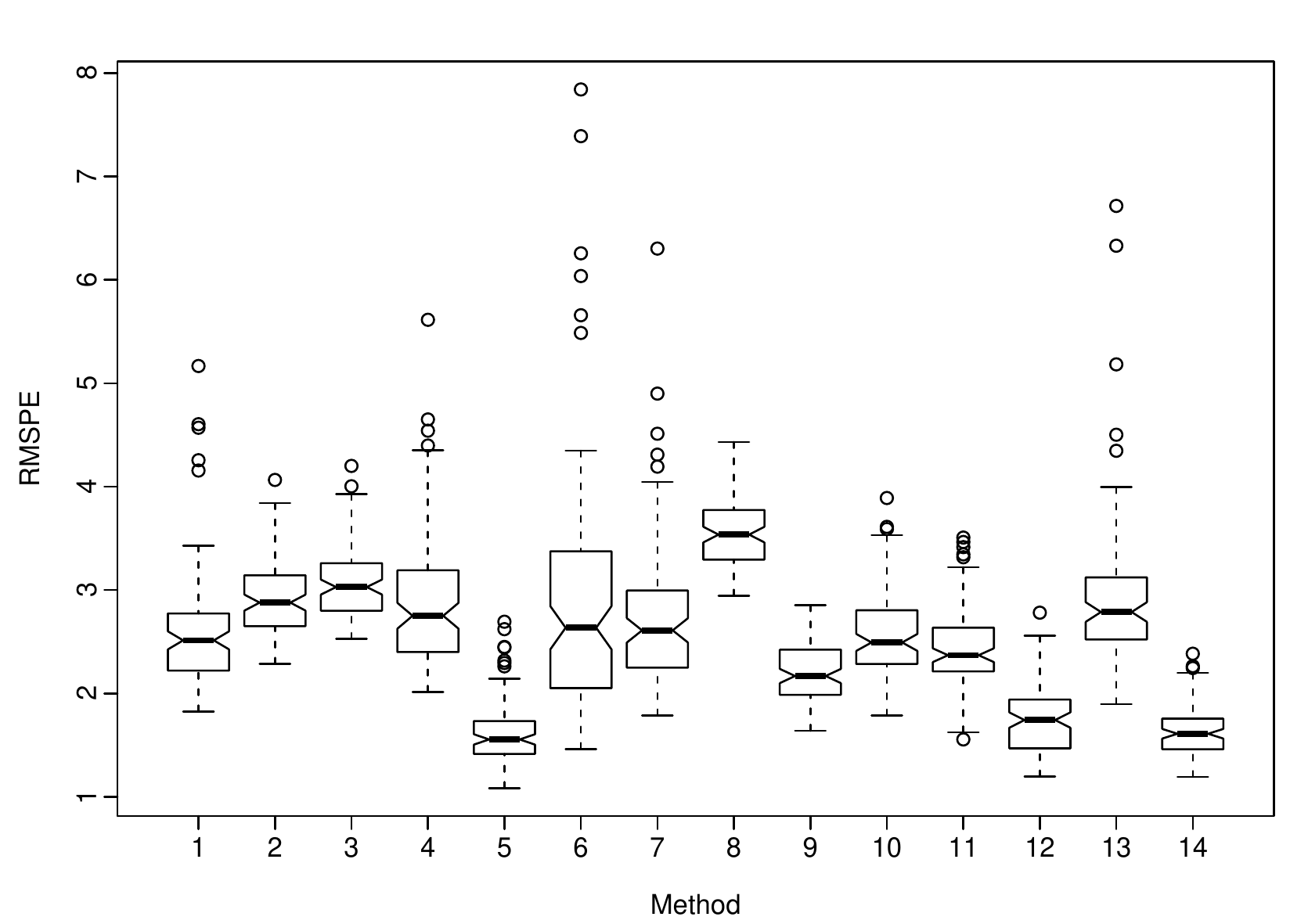}
\caption{Comparisons of forecast accuracy for the 14 functional regression models listed above.}\label{fig:spec_compar}
\end{figure}

The boxplot of RMSPEs is presented in Figure~\ref{fig:spec_compar}. The functional partial linear model with the semi-metric and bandwidth selected by the Bayesian method has the smallest RMSPE among all methods considered. 

\begin{figure}[!ht]
\centering
\includegraphics[width=14cm]{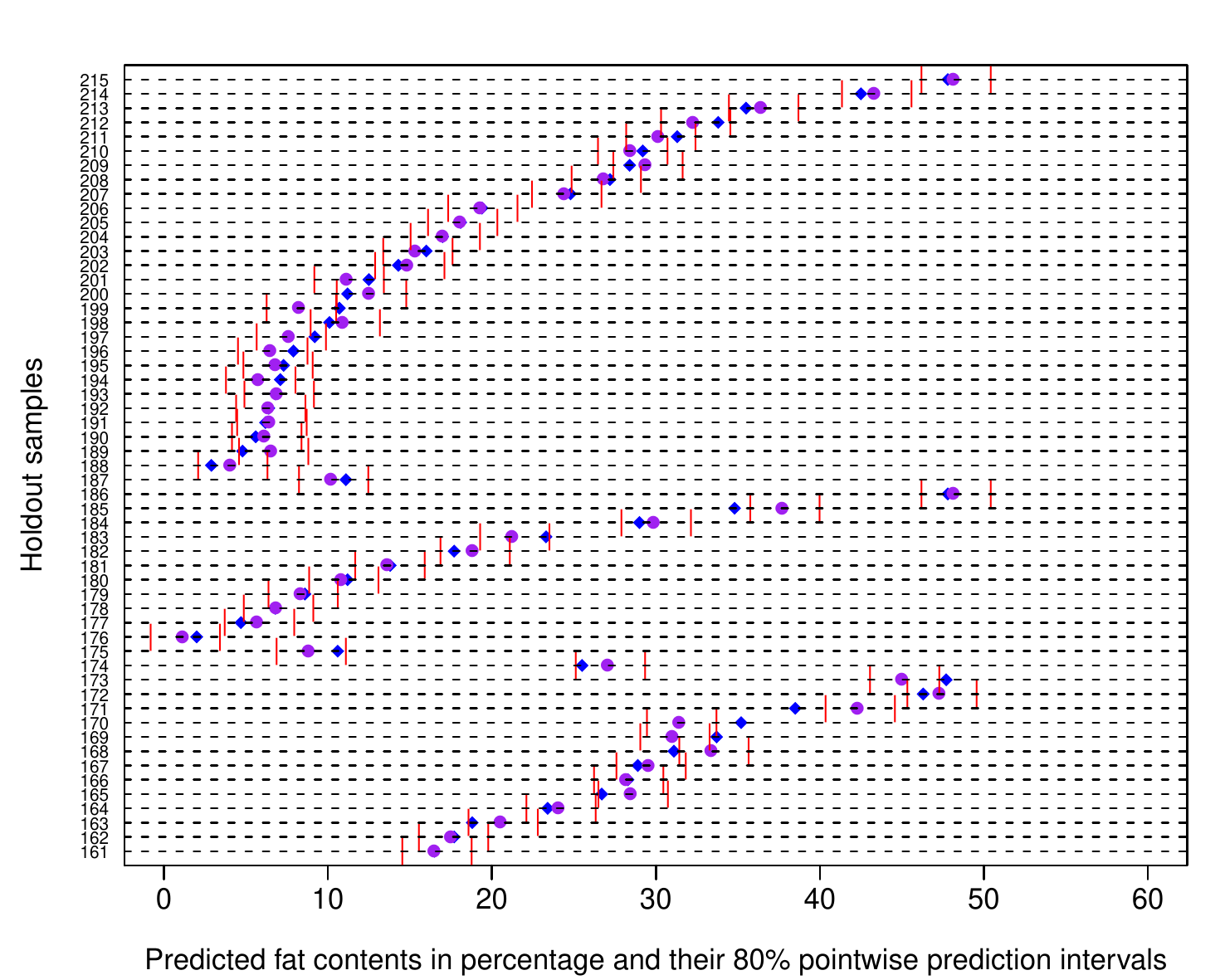}
\caption{A plot of predicted fat content in percentage and their 80\% pointwise prediction intervals. The actual holdout data are presented as diamond-shaped dots; the point forecasts of the fat content are presented as circles; the 80\% pointwise prediction intervals are presented as vertical bars.}\label{fig:CI}
\end{figure}

A by-product of the Bayesian method is that it is possible to compute the pointwise prediction interval nonparametrically. To make this computation, we first compute the cumulative density function of the error density, over a set of grid points within a range, such as between -10 and 10. Then, we take the inverse of the cumulative density function and find two grid points that are closest to the 10\% and 90\% quantiles. The 80\% pointwise prediction interval of a holdout sample is obtained by adding the two grid points to a point forecast. In Figure~\ref{fig:CI}, the holdout fat content in percentage is presented as diamond-shaped dots, the point forecasts of the fat content are presented as round circles, and the 80\% pointwise prediction intervals are presented as vertical bars. At the nominal coverage probability of 80\%, the empirical coverage probability is 87\%; at the nominal coverage probability of 50\%, the empirical coverage probability is 51\%. 

\section{Conclusion}\label{sec:6}

We propose a Bayesian method to estimate optimal bandwidths in a functional partial linear model with homoscedastic errors and unknown error density. As a by-product of the MCMC sampling algorithm, it allows us to compute log marginal likelihood for a semi-metric. The optimal semi-metric is the one that provides the smallest log marginal likelihood, as evident from the results in Table~\ref{tab:3}. Given that no closed-form expression for our bandwidth estimator exists, establishing the mathematical properties, such as the asymptotic optimality of \cite{Shibata81}, of the bandwidth estimator has been difficult. However, we have developed an approximate solution to the bandwidth estimator through MCMC. As a by-product of the Bayesian method, a prediction interval can be obtained, and marginal likelihood can also be used to determine the optimal choice of semi-metric among a set of semi-metrics.

Through a series of simulation studies, we find the functional partial linear model has better estimation and forecast accuracies than functional principal component regression and functional nonparametric regression. For a set of smooth curves, we further confirm the optimality of the semi-metric based on the second derivative; for a set of rough curves, we affirm the overall optimal semi-metric is the semi-metric based on the FPCA, followed closely by the semi-metric based on the first derivative. The kernel density estimator with localized bandwidths performs similarly to the kernel density estimator with a global bandwidth for estimating regression function and outperforms the one with a global bandwidth for estimating error density. 

Using the spectroscopy dataset, the functional partial linear model produces the smallest forecast error of some commonly used functional regression models. The Bayesian method not only allows determination of the optimal semi-metric via marginal likelihood, but it also allows the nonparametric construction of pointwise prediction intervals for measuring prediction uncertainty. 

There are many ways in which the Bayesian method and functional partial linear model can be extended, and we briefly outline five:
\begin{enumerate}
\item[1)] Combine different semi-metrics with weights based on their marginal likelihoods; this leads to the idea of ensemble forecasting.
\item[2)] Consider other functional regression estimators, such as the functional local-linear estimator of \cite{BFR+07} or the $k$-nearest neighbor estimator of \cite{BFV09} -- the functional local-linear estimator can improve the estimation accuracy of the regression function by using a higher-order kernel function; the $k$-nearest neighbor estimator considers the local structure of the data and gives better forecasts when the functional data are heterogeneously concentrated.
\item[3)] Extend to a functional partial linear model with heterogeneous errors; the covariate-dependent variance can be modeled by another kernel-density estimator \citep[e.g.,][]{CCP09}.
\item[4)] Extend to a functional partial linear model with autoregressive errors \citep[e.g.,][]{YZ06}.
\item[5)] Consider a $J$ test of \cite{DM81} for selecting non-nested functional regression models. 
\end{enumerate}

\newpage
\bibliographystyle{apalike}
\bibliography{fplm}
  
\end{document}